\documentclass[a4paper,11pt]{article}
\pdfoutput=1

\usepackage{style} 
\usepackage[T1]{fontenc} 
\usepackage[utf8]{inputenc}
\usepackage{float}
\usepackage{subcaption}
\usepackage{amsmath,amssymb}
\usepackage{graphicx}
\usepackage[colorinlistoftodos]{todonotes}
\title{Spatial entanglement in interacting Bose-Einstein condensates}

\author{N. S\'anchez-Kuntz and}
\author{S. Floerchinger}

\affiliation{Institut f\"ur Theoretische Physik Universit\"at Heidelberg,\\Philosophenweg 16, D-69120 Heidelberg}
\emailAdd{sanchez@thphys.uni-heidelberg.de}
\emailAdd{stefan.floerchinger@thphys.uni-heidelberg.de}

\abstract{The entanglement between spatial regions in an interacting Bose-Einstein condensate is investigated using a quantum field theoretic formalism. Regions that are small compared to the healing length are governed by a non-relativistic quantum field theory in the vacuum limit, and we show that the latter has vanishing entanglement. In the opposite limit of a region that is large compared to the healing length, the entanglement entropy is like in the vacuum of a relativistic theory where the velocity of light is replaced with the velocity of sound and where the inverse healing length provides a natural ultraviolet regularization scale. Besides the von Neumann entanglement entropy, we also calculate R\'enyi entanglement entropies for a one-dimensional quasi-condensate.}

\begin{document}
\maketitle
\flushbottom
\section{Introduction}

In the second half of the 1980's it was shown that the vacuum state of a relativistic quantum field theory violates Bell's inequalities \cite{SuWe1985, SuWe1987, La1987, La1987b}, and subsequently the quantification of entanglement within vacuum states became an important task. Later it was shown that the entanglement entropy associated to a spatial region in a relativistic quantum field theory displays both ultraviolet (UV) and  infrared (IR) divergences \cite{HoLa1994, BePi1996}. 

Initially entanglement in relativistic quantum field theories was investigated mainly with a view on the black hole information paradox \cite{BoKo1986,Sr1993,CaWi1994}, and more recently in the context of holography \cite{RyTa2006,RyTa2006b}. Here we argue that a very similar kind of entanglement is also at play in interacting Bose-Einstein condensates when probed at long distances, where it may be amendable to experimental investigation \cite{IsMa2015,StMu2014,KuPr2018}.

Formally we may start from the R\'{e}nyi entanglement entropy which quantifies entanglement between some region $A$ and its complement region $B$ (such that $A$ and $B$ together form a Cauchy hypersurface of space-time, for example a hypersurface of constant time),
\begin{equation}
S_\alpha(A) = -\frac{1}{\alpha-1}\ln \mbox{Tr}\{\rho_A^\alpha\}.
\label{eq:EE}
\end{equation}
Here $\rho_A=\text{Tr}_B\{\rho\}$ is the reduced density matrix for the region $A$. The von Neumann entanglement entropy is obtained from \eqref{eq:EE} in the limit $\alpha \rightarrow 1$, and can be written as $S(A) = - \mbox{Tr}\{\rho_A\ln\rho_A\}$. It was observed that the leading UV divergence of this quantity is proportional to the area of the boundary that separates regions $A$ and $B$ \cite{Sr1993}, see ref.\ \cite{CaHu2009} for a more detailed discussion and generalization to arbitrary number of space dimensions. 

While it is straight forward to introduce convenient UV and IR regularizations (such as a spatial lattice and finite volume), one is ultimately interested in universal quantities that are independent of the precise regularization scheme. One possibility is to take derivatives of $S_\alpha(A)$, for example with respect to the size of region $A$, until one arrives at a finite result, another is to work with relative entanglement entropies or mutual informations, for which at least the UV divergences cancel out. This is possible because the UV divergent terms are actually independent of the state.

It was realized that conformal transformations  substantially simplify determinations of entanglement entropy on a technical level. Accordingly, a rather detailed understanding of entanglement entropy is now available for conformal field theories \cite{HoLa1994,ViLa2003,Ko2004,CaCa2004,CaCa2009}. For example, the entanglement entropy associated to an interval of length $L$ in a $1+1$ dimensional conformal field theory is given by
\begin{equation}
S(L)=\frac{c}{3} \ln\left(\frac{L}{\epsilon}\right) + \text{const}.
\label{eq:1dentropy}
\end{equation}
In this equation $c$ is the conformal central charge and $\epsilon$ is a small length that regularizes the UV divergences.
The additive constant in \eqref{eq:1dentropy} is not universal and depends on the details of the UV regularization. Moreover, one expects this constant to depend also on the IR regularization and it may even diverge logarithmically when the IR regulator is removed.

The issue of IR divergences is actually an interesting one, and they are so far less studied than the state-independent UV divergences. Infrared divergences arise typically in the presence of gapless excitations, or massless particles in the relativistic jargon, at least in $d=1+1$ dimensions. In ref.\ \cite{Un1990} it was argued that the homogeneous or zero mode is actually responsible for IR divergences because its amplitude is not restricted energetically (see also refs.\ \cite{MaTi2014,ChSh2019,HuKe2017,Ya2017} for related work). Analytic as well as numerical calculations for free theories showed an infrared divergence attributed to this mode \cite{CaHu2009}. In our formalism we find an IR divergence associated to a homogeneous mode, as well. We give an intuitive physical explanation of this phenomenon in section \ref{sec:Phase}. As a consequence of gapless zero modes in the theory, even the relative entropy and mutual information become infrared divergent quantities; this is discussed in refs.\ \cite{Sr1993, CaHu2004, CaHu2009}. In ref.\ \cite{CaHu2009} the infrared divergence of both, the mutual information and the entanglement entropy for a free one-dimensional relativistic scalar theory was found proportional to 
\begin{equation} 
\frac{1}{2} \ln(-\ln(m R)),
\label{eq:IRdivergence}
\end{equation}
where $m$ is a small mass with $m \rightarrow 0$ and $R$ a characteristic length scale.
Besides introducing a small mass $m$ there are other ways to regularize the IR sector, for us it will be convenient to introduce an infrared regulator at the momentum scale $\mu / L$ by hand. A finite temperature also provides a physical IR regulator, as shown in ref.\ \cite{So2011}.

In the main part of this work we will calculate the entanglement entropy of a non-relativistic,  interacting quantum field theory describing bosons with contact interactions, in a Gaussian saddle point approximation to the functional integral. Fluctuations follow the well-known Bogoliubov dispersion relation 
\begin{equation}
\omega = \sqrt{\frac{\mathbf{p}^2}{2M}\left(\frac{\mathbf{p}^2}{2M}+2\lambda \rho\right)},
\label{eq:disrel}
\end{equation}
of which one can find two limits: an effectively relativistic limit for small momenta, with a linear dispersion relation $\omega\approx c_s |\mathbf{p}|$, and a non-relativistic free particle limit for high momenta, $\omega\approx  \frac{\mathbf{p}^2}{2M}$. The fact that in the UV the theory becomes non-relativistic induces a physical UV cutoff to the entanglement entropy also in the ``relativistic regime'' corresponding to large distances $L\gg \xi$. The transition scale, as well as the scale of the natural UV regularization are given by the healing length $\xi$ (defined in eq.\ \eqref{eq:defHealingLength} below). Interestingly, for small interval length $L\ll \xi$ we find vanishing entanglement entropy. This can be understood from the vacuum limit of the non-relativistic quantum field theory, which does not show any entanglement between regions.

Because the low momentum modes behave like a relativistic theory of massless particles we also observe an infrared divergence which we regularize by introducing a lower momentum bound at the scale $\mu/L$ where $\mu$ is a small number. For non-relativistic atoms, this is ultimately also related to the size of the entire condensate, because no real divergence is expected for a finite number of atoms. 
For finite non-relativistic systems the entanglement entropy would be finite in both, the UV and the IR.

In this work we implement the approach developed in ref.\ \cite{BeFlVe2018} based on the symplectic eigenvalues formalism (see also refs.\ \cite{EiCr2010, De2012}), through which one can calculate entanglement for Gaussian states in a QFT, solely in terms of the connected two-point correlation functions. In order to better range our results and compare them to the non-interacting case, we discuss first briefly the entanglement entropy for non-interacting Bose-Einstein condensates in number eigenstates in section \ref{QM}. Subsequently in section \ref{sec:Phase} we perform another warm up exercise by investigating qualitatively entanglement through long-range phase correlations, specifically the homogeneous phonon mode.

In section \ref{entanglement_qft} we then discuss first the field theoretic formalism for the quantification of entanglement, and in section \ref{bogoliubov_theory} we specialize to Bogoliubov theory. This anticipates already some specific results: the entropy of the entire, unbounded system, the non-interacting limit, and the vacuum limit of vanishing condensate.

In section \ref{entropy_one_d} we specialize to an effectively $1+1$ dimensional situation with a quasi-condensate. In section \ref{matrix} we calculate the matrix elements needed for our entropy calculation (this is the main technical work), while section \ref{numerical} is devoted to our numerical results. An analysis of the results is given there, and our conclusions are presented in section \ref{conclusions}. In appendix \ref{ap:modeexpansion} we present technical details about a Fourier expansion scheme with arbitrary boundary conditions that we use for our numerical calculations.

\section{Entanglement in a non-interacting Bose-Einstein condensate}
\label{QM}

As a preparation to the quantum field theoretic discussion for an interacting Bose-Einstein condensate let us recall here the entanglement properties of finite,  non-interacting Bose-Einstein condensates at fixed particle number, which can be described in a quantum mechanical formalism \cite{Si2002}.

The basic idea can be formulated in rather general terms. We consider a situation with appropriate boundary conditions such that the single particle Hamiltonian has a unique ground state with homogeneous amplitude. An example would be a box with periodic boundary conditions and a unique zero mode. A many-body state with $N$ non-interacting atoms has a ground state where all of them occupy the zero mode. 

Now let us split the volume of the box into two parts, $A$ and $B$, with relative volumes $w=V_A/V$ and $1-w=V_B/V$. For a single particle one has now a probability $w$ to find it in the region $A$ and $1-w$ to find it in $B$. For $N$ particles one expects a binomial distribution of the occupation numbers. The $N$-particle state can therefore be written as
\begin{equation}
| \psi \rangle = \sum_{k=0}^N \sqrt{\binom{N}{k} w^k (1-w)^{N-k}} \;e^{i\alpha_k}  \; | k \rangle_A | N-k \rangle_B
\end{equation}
where $|k \rangle_A$ denotes a state with $k$ particles in a homogeneous state in the region $A$ and similarly $| N-k \rangle_B$ for region $B$. The phases $\alpha_k$ are not determined at this point. We assume a normalization $\langle m| n \rangle = \delta_{mn}$ for both subsystems. 

One may now consider the density matrix for the entire state $\rho = | \psi \rangle \langle \psi |$, as well as the reduced density matrix for subsystem $A$,
\begin{equation}
\rho_A = \text{Tr}_B \{ \rho \} = \sum_{k=0}^N \binom{N}{k} w^k (1-w)^{N-k} |k \rangle \langle k |,
\label{eq:mixedDensityMatrixRhoA}
\end{equation}
where we have dropped the index $A$ on the right hand side for convenience. We observe that $\rho_A$ describes now a mixed state, except in the limiting cases $w=1$, $w=0$, and $N=0$. This limiting cases correspond either to $V_A=V$, $V\rightarrow \infty$, or a vanishing condensate. Otherwise, the particle number in the region $A$ is fluctuating, and the local state is a mixed superposition of the different possibilities. 

Because \eqref{eq:mixedDensityMatrixRhoA} is diagonal, one can immediately determine the corresponding von Neumann entropy as the information entropy of a binomial distribution. For large $N$ the latter can be written as
\begin{equation}
S(\rho_A) = \frac{1}{2} \ln\left( 2\pi e N w (1-w) \right) + \mathcal{O}(1/N).
\end{equation}

A limiting case is that of a very large particle number $N\to \infty$ and a small volume $V_A$, where $w\to 0$ in such a way that the expected particle number in region $A$ given by $M=w N$ remains finite. In that case the binomial distribution approaches a Poisson distribution,
\begin{equation}
\binom{N}{k} w^k (1-w)^{N-k} \to \frac{M^k}{k!} e^{-M}.
\end{equation}
For $M\gg 1$, the entropy of the Poisson distribution can then be approximated as
\begin{equation}
S(\rho_A) = \frac{1}{2} \ln\left( 2\pi e M \right) - \frac{1}{12 M} - \frac{1}{24 M^2}- \frac{19}{360 M^3}+ \mathcal{O}(1/M^4).
\end{equation}
More generally, the entropy of the Poisson distribution leads to the expression
\begin{equation}
\begin{split}
S(\rho_A) & = M \left[ 1- \ln(M) \right] + e^{-M} \sum_{k=0}^\infty \frac{M^k \ln(k!)}{k!} \\
& = M \left[ 1- \ln(M) \right] + \frac{\ln(2)}{2}M^2 - \frac{\ln(4/3)}{6} M^3 + \mathcal{O}(M^4).
\end{split}
\label{eq:qmentropy}
\end{equation}
Note that the power series on the right hand side in the first line has infinite radius of convergence. For $M\ll 1$ it is dominated by the first few terms as written out in the second line. 

We observe that this result has still a dependence on the particle number in the region $A$ but does otherwise not depend on $w$ any more. For $M\to 0$ one finds $S(\rho_A) \to 0$ as expected.

For states with a fixed number of non-interacting particles in the ground state, the entanglement is essentially an entanglement of particle number. Indeed, if one measures particle number in subsystem A as $k$, one can immediately infer the particle number in the complement subsystem B as $N-k$. One can actually extend the above considerations to states where particle number is fluctuating, such as coherent states \cite{Si2002}. A coherent state describing a non-interacting Bose-Einstein condensate at non-zero chemical potential, but zero temperature, shows then no entanglement between spatial regions at all. This is in agreement with the fact that no information can be gained from measuring particle number locally in a subregion. We will see below that things change again in the presence of interactions and that the ground state of an interacting Bose-Einstein condensate shows sizable entanglement between regions.

\section{Phase entanglement through the homogeneous mode}
\label{sec:Phase}

As a further preparatory step for the field theoretic discussion of entanglement entropy below, it is useful to consider another simplified model. Here we consider a Bose-Einstein condensate of infinite extend at finite density and we are specifically interested in the homogeneous or zero wave number phonon mode. Let us anticipate the following action, which describes Bogoliubov phonons in the low momentum or long distance regime,
\begin{equation}
S = \int d^4 x \left\{ \frac{1}{2} (\partial_t \chi)^2 - \frac{1}{2}c_s^2 (\vec \nabla \chi)^2  \right\}.
\label{eq:actionPhonons}
\end{equation}
The field $\chi = \varphi_2 / \sqrt{2\lambda \rho}$ is essentially the phase of the order parameter field, see eq.\ \eqref{eq:fieldDecomposition} for the precise definition of $\varphi_2$. Eq.\ \eqref{eq:actionPhonons} describes phonons propagating with the velocity of sound $c_s$ as defined in eq.\ \eqref{eq:velocityOfSound}. The Hamiltonian corresponding to \eqref{eq:actionPhonons} can be written as
\begin{equation}
H = \int \frac{d^3 p}{(2\pi)^3} \left\{ \frac{1}{2} \pi(-\mathbf{p}) \pi(\mathbf{p}) + \frac{1}{2} \chi(-\mathbf{p}) \, c_s^2 \, \mathbf{p}^2 \, \chi(\mathbf{p}) \right\},
\end{equation}
where $\pi = \partial_t \chi$ is the conjugate momentum of $\chi$. While modes with $\mathbf{p}^2>0$ have a Hamiltonian (and therefore ground state) similar to the quantum mechanical harmonic oscillator, the situation is different for the homogeneous zero mode with $\mathbf{p}=0$. The latter has a Hamiltonian as a free particle in quantum mechanics, $H = \text{const} \times \pi^2$. This implies that the phase field $\chi$ is not constrained in the ground state; it can have any (time-independent) value or superposition thereof. This is directly related to the phenomenon of spontaneous symmetry breaking and long-range order.

Let us now discuss what this implies from an entanglement point of view. Because the homogeneous part of $\chi$ is unconstrained by energy minimization, the vacuum state can be formally in a superposition of different values. Such a situation can lead to a large amount of entanglement between regions. While the homogeneous part of the phase is \textit{a priori} in a superposition, it is fixed everywhere once it is measured in one region. In this sense, different regions are therefore entangled. Moreover, because $\chi$ is a continuous degree of freedom, the corresponding entanglement entropy can become formally infinite.

We have identified here a physical reason for the infrared divergence in the entanglement entropy which we will also observe in our field theoretic calculations below. Note that this phenomenon would be absent in the presence of a small energy gap, for example when one replaces $c_s^2 \mathbf{p}^2 \to c_s^2 \mathbf{p}^2 + m^2$.
For a field theoretic calculation of entanglement entropy also the modes in the vicinity of $\mathbf{p}=0$ are expected to be important. More detailed calculations are needed to show for which theories, and for which states, infrared divergences actually occur.

We should also remark here that in a realistic physics situation there might be strong decoherence effects that destroy the macroscopic entanglement associated to the homogeneous part of the phase field $\chi$. It should then be treated as a classical variable, the value of which is chosen by spontaneous symmetry breaking as in a classical treatment of the phenomenon.

\section{Spatial entanglement of quantum fields}
\label{entanglement_qft}

We now wish to investigate spatial entanglement in an interacting Bose-Einstein condensate from a quantum field theoretic point of view. In a quantum field theory, the degrees of freedom are of course the fields themselves. The density matrix for the state at some time $t$ becomes a functional $\rho[\phi_+,\phi_-]$. One may define a projection operator $\mathsf{P}$ such that $\mathsf{P}\phi(x)=\phi(x)$ for positions $x$ in the region $A$ and  $\mathsf{P}\phi(x)=0$ for positions $x$ in the complement region $B$. The reduced density matrix for the region $A$ reads then 
\begin{equation}
\rho_A [\phi_+,\phi_-] = \frac{1}{Z_B} \int \mathcal{D} \tilde{\phi}\, \rho[\mathsf{P}\phi_+ + (1-\mathsf{P})\tilde\phi \, , \,\mathsf{P}\phi_- + (1-\mathsf{P})\tilde\phi]
\end{equation}
where $Z_B$ is chosen such that $\mbox{Tr}\{\rho_A\} = 1$.

The von Neumann entanglement entropy would then be calculated through $S_A = -\mbox{Tr}\{\rho_A \mbox{ln}\rho_A\}$. Typically, a concrete determination of the entanglement entropy for a quantum field theory in a generic state $\rho$ is difficult. Some simplification takes place when $\rho[\phi_+,\phi_-]$ is a Gaussian density matrix, i.e. when $\mbox{ln}(\rho[\phi_+,\phi_-])$ is a quadratic functional of the fields $\phi_+$ and $\phi_-$. For an interacting Bose-Einstein condensate, the equilibrium state is not exactly Gaussian, but a Gaussian approximation is expected to be close to reality. Many of the equilibrium properties can be well described by Bogoliubov theory in terms of approximately free Bogoliubov quasi-particles. This is in particular a good approximation when the interaction strength is not too strong. In the following we will investigate the entanglement entropy within the Bogoliubov approximation, which corresponds to a Gaussian density matrix.

As one may expect, in the case of Gaussian states, the entropy depends only on connected two-point correlation functions. This situation, and the corresponding mathematical formalism, are discussed in detail in ref. \cite{BeFlVe2018}. For the present paper we can restrict to a particular situation, namely where the mixed statistical correlation function of fields $\phi$ and their conjugate momenta $\pi$ vanishes, $\langle \phi\pi + \pi\phi \rangle = 0$. In this case, the R\'enyi entropy can be expressed through
\begin{equation}
S_\alpha =\frac{1}{(\alpha-1)} \left\{\mbox{Tr} \ln \left(\left(\sqrt{a + \frac{1}{4}} + \frac{1}{2}\right)^\alpha - \left(\sqrt{a + \frac{1}{4}} - \frac{1}{2}\right)\right)^\alpha\right\}.
\label{eq:Rentropy}
\end{equation}
One can take the limit $\alpha\rightarrow 1$ of the above expression, and see that the von Neumann entropy is obtained through
\begin{equation}
\begin{split}
S=& \, \mbox{Tr} \left\{\left(\sqrt{a + \frac{1}{4}} + \frac{1}{2}\right) \ln \left(\sqrt{a + \frac{1}{4}} + \frac{1}{2}\right) - \left(\sqrt{a + \frac{1}{4}} - \frac{1}{2}\right) \ln \left(\sqrt{a + \frac{1}{4}} - \frac{1}{2}\right)\right\}.
\end{split}
\label{eq:vNentropy}
\end{equation}
The matrix $a$ is given in position representation by
\begin{equation}
a(t, \mathbf{x},\mathbf{y}) = \int_{\mathbf{z}} \langle \phi (t, \mathbf{x}) \phi (t, \mathbf{z})\rangle \langle \pi (t, \mathbf{z}) \pi (t, \mathbf{y})\rangle - \frac{1}{4} \delta(\mathbf{x}-\mathbf{y}),
\label{eq:matrixa}
\end{equation}
where the correlation functions correspond to the statistical equal time correlation functions (see below for a more detailed discussion). A very interesting feature of eq. \eqref{eq:vNentropy} is that it holds both, for the global von Neumann entropy as well as for entanglement entropies. The only difference in the latter case is that the matrix or operator trace in \eqref{eq:vNentropy} and the integral over positions in \eqref{eq:matrixa} need to be restricted accordingly.

\section{Non-relativistic bosonic quantum field theory}
\label{bogoliubov_theory}
We now consider the following microscopic action for a complex non-relativistic scalar field $\varphi(x)$,  

\begin{equation}
S[\varphi] = \int  d^4x  \left\{ \varphi^* \left(i \partial_t + \frac{\nabla^2}{2M} - V_0 \right)\varphi  - \frac{\lambda}{2}(\varphi^* \varphi)^2 \right\}.
\label{eq:nonrelaction}
\end{equation}
Here $V_0(x)$ is an external potential and $\lambda$ is a (bare) contact interaction parameter. At finite density $V_0$ can be replaced by the chemical potential, $V_0 \to - \mu$.

While \eqref{eq:nonrelaction} represents an interacting quantum field theory that cannot be solved exactly, a good qualitative, and at sufficiently small coupling $\lambda$ also quantitative understanding can be gained from a Gaussian or steepest descent approximation. First, the action \eqref{eq:nonrelaction} has a stationary point with homogeneous background field $\varphi(x) = \phi_0$ when 
\begin{equation}
\rho = \phi_0^*\phi_0=\frac{\mu}{\lambda}.
\end{equation}
Furthermore, if one considers perturbations to this  background field, 
\begin{equation}
\varphi= \phi_0 + [\varphi_1 + i\varphi_2]/\sqrt{2}, 
\label{eq:fieldDecomposition}
\end{equation}
one obtains for the action to quadratic order in these perturbations,  
\begin{equation}
S[\varphi_1,\varphi_2] =  \int  d^4x\left\{ \mu \rho + \frac{\lambda}{2} \rho^2 - \frac{1}{2} (\varphi_1,\varphi_2) \left(\begin{array}{cc}-\frac{\nabla^2}{2M}+2\lambda \rho \, & \, \partial_t\\-\partial_t&\,-\frac{\nabla^2}{2M}\end{array}\right) \binom{\varphi_1}{\varphi_2} \right\}.
\label{eq:qactperturbations}
\end{equation}
When written in momentum space, eq.\ \eqref{eq:qactperturbations} directly yields the inverse propagator for Bogoliubov theory, so that the propagator reads
\begin{equation}
G(p)= \frac{1}{- \omega^2 + \frac{\mathbf{p}^2}{2M} \left(\frac{\mathbf{p}^2 }{2M}+ 2 \lambda \rho\right)} 
\left(\begin{array}{cc} 
\frac{\mathbf{p}^2}{2M}  & i\omega \\
- i\omega \, & \, (\frac{\mathbf{p}^2}{2M}+ 2 \lambda \rho)
\end{array}\right).
\label{eq:propagator}
\end{equation}
The propagator has poles on the Bogoliubov dispersion relation \eqref{eq:disrel}. A characteristic length scale is given by the {\it healing length},
\begin{equation}
\xi = \frac{1}{\sqrt{2M\lambda \rho}}.
\label{eq:defHealingLength}
\end{equation}
For momenta that are small compared to the inverse healing length $|\mathbf{p}| \ll 1/\xi$ the dispersion relation is linear, $\omega \approx c_s |\mathbf{p}|$, while it becomes quadratic, $\omega \approx \frac{\mathbf{p}^2}{2M}$ for large momenta $|\mathbf{p}| \gg 1/\xi$. Note also that in the low momentum regime, corresponding to the long-distance regime for the propagator, we have a dispersion relation as for massless relativistic particles, but involving instead of the speed of light the {\it velocity of sound}
\begin{equation}
c_s=\sqrt{\frac{\lambda \rho}{M}}.
\label{eq:velocityOfSound}
\end{equation}

Furthermore, from \eqref{eq:qactperturbations} we obtain the conjugate momenta of the field $\varphi_1$ and $\varphi_2$, respectively,
\begin{equation}
\pi_1(x) = \frac{\delta S}{\delta {\dot{\varphi}_1(x)}} = \varphi_2(x), \quad\quad\quad \pi_2(x) = \frac{\delta S}{\delta \dot{\varphi}_2(x)} = -\varphi_1(x).
\end{equation}
This shows that in the low-momentum or long distance regime one can expect the theory to be equivalent to a relativistic theory with a single real massless scalar field $\phi=\varphi_1$, while $\varphi_2$ is simply its conjugate momentum.

\subsection{Spectral density and statistical correlation functions}
\label{spectral}

It is also convenient to introduce the spectral density $\rho_{ij}(\omega, \mathbf{p})$ for the different field components such that the propagator can be written as
\begin{equation}
G_{ij}(p^0, \mathbf{p})=  \int_{-\infty}^{\infty} d\omega \frac{\rho_{ij}(\omega, \mathbf{p})}{\omega - p^0},
\label{eq:tospectden}
\end{equation}
from which one can determine the spectral correlation function given by (we use the notation of ref.\ \cite{Fl2016})
\begin{equation}
\Delta_{ij}^\rho (p) = 2 \pi \rho_{ij} (p^0, \mathbf{p}).
\label{eq:tospectcorr}
\end{equation}

By substituting in \eqref{eq:tospectden} with the diagonal elements of Green's function \eqref{eq:propagator} one obtains the diagonal components of the spectral density matrix,
\begin{equation}
\begin{split}
\rho_{\phi \phi} (\omega, \mathbf{p})
=& \frac{1}{2}\sqrt{\frac{\mathbf{p}^2}{ \mathbf{p}^2 + 4M \lambda \rho }} \left[ \delta \left(\omega - \sqrt{\tfrac{\mathbf{p}^2 }{2M} \left( \tfrac{\mathbf{p}^2 }{2M} + 2 \lambda \rho \right)}\right)
-  \delta \left( \omega + \sqrt{\tfrac{\mathbf{p}^2 }{2M} \left( \tfrac{\mathbf{p}^2 }{2M} + 2 \lambda \rho \right) }\right)    \right] 
\end{split}
\end{equation}
and
\begin{equation}
\begin{split}
\rho_{\pi \pi} (\omega, \mathbf{p}) 
=& \frac{1}{2} \sqrt{\frac{\mathbf{p}^2 + 4M \lambda \rho }{\mathbf{p}^2}} \left[ \delta \left( \omega - \sqrt{\tfrac{\mathbf{p}^2 }{2M} \left( \tfrac{\mathbf{p}^2 }{2M} + 2 \lambda \rho \right) }\right) 
-  \delta \left( \omega + \sqrt{\tfrac{\mathbf{p}^2 }{2M} \left( \tfrac{\mathbf{p}^2 }{2M} + 2 \lambda \rho \right) }\right)  \right].
\end{split}
\end{equation}
In addition to this, we have the off-diagonal components
\begin{equation}
\rho_{\phi \pi} (\omega, \mathbf{p}) = -\rho_{\pi\phi} (\omega, \mathbf{p}) = \frac{i}{2} \left[ \delta \left( \omega - \sqrt{\tfrac{\mathbf{p}^2 }{2M} \left( \tfrac{\mathbf{p}^2 }{2M} + 2 \lambda \rho \right) }\right)
+  \delta \left( \omega + \sqrt{\tfrac{\mathbf{p}^2 }{2M} \left( \tfrac{\mathbf{p}^2 }{2M} + 2 \lambda \rho \right) }\right)  \right].
\end{equation}

Given the spectral densities, one can immediately determine various versions of Green's functions, such as retarded, advanced, time-ordered and anti-time-ordered, see e.\ g.\ ref.\ \cite{Fl2016}. Moreover, in thermal equilibrium, statistical correlation functions $\Delta_{ij}^S (p)$ are related to the spectral correlation functions $\Delta_{ij}^\rho (p)$ through the fluctuation-dissipation relation,
\begin{equation}
\Delta_{ij}^S (p) = \left[ \frac{1}{2} + n_\text{B}(p^0) \right] \Delta_{ij}^\rho (p),
\label{eq:FluctDissRel}
\end{equation}
with the Bose-Einstein thermal distribution function $n_\text{B}(\omega) = 1/(e^{\omega/T}-1)$. Note that the square bracket on the right hand side of \eqref{eq:FluctDissRel} is anti-symmetric under $\omega\to -\omega$.

Specifically we find the equal-time statistical correlation functions
\begin{equation}
	\begin{split}
 \Delta_{\phi \phi}^S(\mathbf{x}-\mathbf{y}) =&\, \int_{\mathbf{p}} \left[ \frac{1}{2} + n(\mathbf{p}) \right]\sqrt{\frac{\mathbf{p}^2}{ \mathbf{p}^2 + 4M \lambda \rho }}e^{i\mathbf{p}(\mathbf{x}-\mathbf{y})}, \\
\Delta_{\pi \pi}^S (\mathbf{x}-\mathbf{y}) =&\, \int_{\mathbf{p}} \left[ \frac{1}{2} + n(\mathbf{p}) \right] \sqrt{\frac{\mathbf{p}^2 + 4M \lambda \rho }{ \mathbf{p}^2 }} e^{i\mathbf{p}(\mathbf{x}-\mathbf{y})},
	\end{split}
\label{eq:statistical}
\end{equation}
where $n(\mathbf{p})$ is now the Bose-Einstein distribution evaluated on the dispersion relation \eqref{eq:disrel}. The mixed correlation functions $\Delta_{\phi \pi}^S$ and $\Delta_{\pi\phi}^S$ vanish at equal times because the corresponding spectral density is symmetric under $\omega \to - \omega$. 

\subsection{Entanglement entropy for three limiting cases} 
\label{NRL}

\paragraph{Homogeneous Bose-Einstein condensate without boundaries.}

As a preview exercise it is interesting to calculate the entropy of the complete system, i.e., a Bose-Einstein condensate at $T=0$, in the ground state with no boundaries. In this case one has vanishing occupation number for phonons, $n(\mathbf{p})=0$, and the matrix $a$ is given by
\begin{equation}
\begin{split}
	a(\mathbf{x},\mathbf{y}) =& \int_{\mathbf{z}}\Delta^S_{\phi \phi}(\mathbf{x},\mathbf{z}) \Delta^S_{\pi \pi}(\mathbf{z},\mathbf{y}) - \frac{1}{4}\delta(\mathbf{x}-\mathbf{y})\, \\
	=& \,\int_{\mathbf{p}}  e^{i\mathbf{p}(\mathbf{x}-\mathbf{y})}\Delta^S_{\phi \phi}(\mathbf{p})\Delta^S_{\pi \pi}(\mathbf{p}) - \frac{1}{4}\delta(\mathbf{x}-\mathbf{y}) = 0,
\end{split}
\end{equation}
which enforces a vanishing entropy $S=0$. Allowing a non-vanishing occupation number for phonons such that $n(\mathbf{p}) \neq 0$, one finds instead the corresponding entropy for a free gas of quasi-particles. If one considers instead an entanglement entropy for some subregion $A$, the correlation functions $\Delta^S_{\phi \phi}(\mathbf{x},\mathbf{y})$ and $\Delta^S_{\pi \pi}(\mathbf{x},\mathbf{y})$ do not change, but the position space integrals need to be restricted to the region $A$. This will lead to a non-vanishing result even at $T=0$.

\paragraph{Non-interacting Bose-Einstein condensate in a coherent state.} 

Another interesting limit is the entanglement entropy of a Bose-Einstein condensate in a coherent state at $T=0$ for vanishing interaction $\lambda=0$. The statistical equal-time correlation functions in \eqref{eq:statistical} simplify to
\begin{equation}
\Delta_{\phi \phi}^{S} (\mathbf{x}-\mathbf{y}) = \Delta_{\pi \pi}^{S} (\mathbf{x}-\mathbf{y}) = \frac{1}{2}\delta(\mathbf{x}-\mathbf{y}),
\label{eq:freecoherent}
\end{equation}
so that the matrix $a$ results in
\begin{equation}
	a(\mathbf{x},\mathbf{y}) = \int_{\mathbf{z}\in A}\Delta^S_{\phi \phi}(\mathbf{x},\mathbf{z}) \Delta^S_{\pi \pi}(\mathbf{z},\mathbf{y}) - \frac{1}{4}\delta(\mathbf{x}-\mathbf{y}) = 0.
	\label{eq:nonrelnonint}
\end{equation}
This leads to vanishing entanglement entropy for any choice of the region $A$. This result has been obtained by different means in ref. \cite{Si2002}. Note that the conclusion would be different for a number eigenstate in a finite volume as discussed in section \ref{QM}.

\paragraph{Fluctuations on top of the vacuum.}
Another interesting limit to study is the one of a non-relativistic quantum field theory in vacuum, i.\ e.\ without any particles. In that case one has $\rho=\phi_0^*\phi_0=0$ and the statistical correlation functions of the fluctuating fields are again of the form \eqref{eq:freecoherent} implying a vanishing entanglement entropy for any choice of the region $A$. This is an interesting result: while the entanglement entropy is typically ultraviolet divergent for a relativistic quantum field theory, it vanishes for a non-relativistic quantum field theory in the vacuum limit. In the non-relativistic theory, all non-vanishing contributions to the entanglement entropy must be due to a non-vanishing number of particles.

Another interesting conclusion can be drawn from this finding: in a non-relativistic quantum field theory the entanglement entropy does not have any ultraviolet divergences, at least at Gaussian level. The reason is here that for very large spatial momenta the dispersion relation always becomes $\omega(\mathbf{p})\rightarrow \mathbf{p}^2/2m$ as in the vacuum state, and the latter is not entangled.

\section{Entanglement entropy in an effectively one-dimensional situation}
\label{entropy_one_d}

Up until now no assumptions were made regarding the shape of region $A$ or even the number of space dimensions. In any case one could determine the quantum field theoretic entanglement entropy for a Gaussian state from eqs.\ \eqref{eq:vNentropy} and \eqref{eq:matrixa} with the appropriate choice of correlation functions and operator traces. To do such calculations in practice can be technically challenging because to compute the logarithm appearing in eq.\ \eqref{eq:vNentropy}, the matrix $a$ must be diagonal, and traces in functional spaces can be also difficult to take. One strategy would be to work with a lattice regularization, as has been done for relativistic field theories \cite{CaHu2009}. In order to obtain the appropriate correlation functions one typically discretizes the entire theory on a spatial lattice and eventually takes the infinite volume and continuum limit numerically.

Here we follow a somewhat different strategy and aim at a calculation of the operator traces directly for the region $A$. To this end we introduce an appropriate mode expansion for this finite region. The advantage is then that this expansion can be truncated for numerical calculations at a large enough wave number (corresponding to a short distance regularization) and the remaining calculations can be done numerically in the resulting finite space. The concrete design of the mode expansion scheme is not always straightforward. In particular, the fields in the region $A$ do not fulfill definite boundary conditions, and accordingly no such boundary conditions should be assumed for the mode expansion. We discuss a specific scheme for a one-dimensional interval in appendix \ref{ap:modeexpansion}.

For higher dimensional situations such as a three-dimensional cube, one could use an adaptation of the scheme in appendix \ref{ap:modeexpansion}. The numerical effort would increase because one would need an appropriate quantum number for each of the spatial dimensions. Another interesting situation would be one where the region $A$ corresponds to a ball of some radius $R$.
Here it would be convenient to work with spherical coordinates and to use angular harmonics for the two angles and an adaptation of the basis in appendix \ref{ap:modeexpansion} for the radial direction. As a consequence of $\mbox{SO}(3)$ rotation symmetry, the two-point correlation functions that enter \eqref{eq:matrixa} can depend only on coordinate differences on the unit sphere $S^2$, and in the appropriate Fourier representation $a$ would be diagonal with respect to the quantum numbers $l,m$ conjugate to angles. However, the correlation function would still be non-diagonal with respect to the quantum numbers conjugate to radius $r$. 

For all possible shapes of the region $A$, the entanglement depends also on the characteristic size $L$ of the region. The general expectation is that for large size $L \gg \xi$ compared to the healing length $\xi$ one has an entanglement entropy in an interacting Bose-Einstein condensate as in a relativistic quantum field theory, but with an ultraviolet momentum regulator given by $1/\xi$. In the present work we concentrate on establishing this numerically and in further detail for a one-dimensional interval, region $A$, while we leave further studies of more complex and higher dimensional regions for the future. 

While restricting the considerations in sections \ref{bogoliubov_theory} and \ref{entropy_one_d} to only one spatial dimension of infinite extend leads to technical simplifications, it has the conceptual drawback that real Bose-Einstein condensation does not exist there. Indeed, the more important role played by (hydrodynamic) fluctuations in lower dimensions destroys proper long-range order \cite{MeWa1966,Ho1967}. In a strictly one-dimensional situation, the Bogoliubov type theory introduced in section \ref{bogoliubov_theory} is therefore not valid. However, the low-momentum excitations are still sound excitations or phonons \cite{LiLi1963} with linear dispersion relation, while at very high momentum it must become quadratic. Furthermore, even though the order parameter might be vanishing for a one-dimensional scenario, it is shown in \cite{LiLi1963} that Bogoliubov's approximation to the dispersion relation is valid for a weakly interacting, high density system. In the intermediate region, however, the dynamic structure function differs substantially from the Bogoliubov result (see e.\ g.\ ref.\  \cite{PitStr2003}).

Given these remarks, in the following we will determine the entanglement entropy for a one-dimensional quasi-condensate within the Bogoliubov approximation described in section \ref{bogoliubov_theory}. On the one hand, this is a preparation for an investigation of more general regions in higher dimensional setups, and on the other hand, also an interesting approximation to which more elaborate calculations of the one-dimensional interacting Bose gas can be compared to. For the $\alpha=2$ R\'enyi entropy a numerical calculation of the entanglement entropy within the Lieb-Liniger model is already available \cite{HeRo2016}. As we will discuss below, our calculations agree rather well with the results reported in ref. \cite{HeRo2016}, which can be seen as an \emph{a posteriori} justification for the Bogoliubov approximation.

In the following we specialize therefore to a one-dimensional space and study how a subregion $A$ (an interval) of length $L$ is entangled with its complement, the infinite space of the real axis without the region $A$. On the technical level, we work with operators (or generalized matrices) $a(x,y)$ where $x,y \in A$ and we will represent these operators in a specifically designed expansion basis, as described in appendix \ref{ap:modeexpansion}.

\subsection{Matrix $a_{mn}$}
\label{matrix}

In the following we calculate the entanglement entropy for the vacuum state within Bogoliubov theory. The starting point is the correlation function in eq.\ \eqref{eq:matrixa} which we will determine in the form $a_{mn}$, where $m$ and $n$ are mode indices according to the expansion introduced in appendix \ref{ap:modeexpansion}. While details are given there, let us anticipate here that the scheme expands functions $f(x)$ on the interval $x \in (0,L)$ as

\begin{equation}
f(x)=\frac{1}{L}\left[f_{-1} + f_{0} \frac{2x-L}{L} + \sum_{n=1}^{\infty} f_n \sin \left( \frac{n\pi}{L}x\right) \right],
\label{eq:expansion}
\end{equation}   
see eq.\ \eqref{eq:nexpL}.

It is convenient to start out with the momentum representation of the equal time statistical correlation functions given by eq.\ \eqref{eq:statistical} (we concentrate on the ground state where $n(p)=0$). To do so we translate these correlation functions to $m,n$-space by making use of the kernels $\tilde{t}_n(p)$ and  $\tilde{s}_n(p)$ obtained in appendix \ref{ap:modeexpansion}. This means that the matrix elements $a_{mn}$ can be written as (we use the abbreviation $\int_p=\int_{-\infty}^{\infty} dp/2\pi$)
\begin{equation}
\begin{split}
a_{mn} =& \, \sum_{l=-1}^{\infty}  [\Delta^S_{\phi \phi}]_{ml} [\Delta^S_{\pi \pi}]_{ln}  - \frac{1}{4} \delta_{mn}  \\
=& \, \sum_{l=-1}^{\infty}  \int_p\int_q \tilde{s}_{m}(p) \Delta^S_{\phi \phi}(p) \tilde{t}_{l}(p) \tilde{s}_{l}(q) \Delta^S_{\pi \pi}(q) \tilde{t}_{n}(q) - \frac{1}{4} \delta_{mn}  \\
=& \, \int_p\int_q \tilde{s}_{m}(p) \frac{1}{2}\sqrt{\frac{p^2}{p^2 + 2/\xi^2}} \mathsf{P}_L(p,q) \frac{1}{2}\sqrt{\frac{q^2 + 2/\xi^2}{q^2}} \tilde{t}_{n} (q) - \frac{1}{4} \delta_{mn},
\label{eq:amnL}
\end{split}
\end{equation}
using a projector to the region $A$ in momentum space, $\mathsf{P}_L(p,q)$ as defined in the last step of eq. \eqref{eq:sumnpq}. We can explicitly calculate the entries of \eqref{eq:amnL} by integrating first over $q$
\begin{equation}
	\begin{split}
I_0(p) =& \, \frac{1}{2}\int_q   \mathsf{P}_L(p,q) \sqrt{\frac{q^2 + 2/\xi^2}{q^2}} \tilde{t}_{n} (q) \\
=& \, \frac{1}{2}\int_q  \frac{e^{i(p- q)(L+\epsilon)}-e^{-i  (p - q)\epsilon}}{i(p-q)} \sqrt{\frac{q^2 + 2/\xi^2}{q^2}} \tilde{t}_{n} (q).
\label{eq:intq}
\end{split}
\end{equation}
This has no poles on the real axis, so that we can slide the contour slightly below and integrate 
\begin{equation}
I_0^a(p) = \frac{1}{2}\int_q  \frac{e^{-i (p - q)\epsilon}}{i(q-p)} \sqrt{\frac{q^2 + 2/\xi^2}{q^2}} \tilde{t}_{n} (q),
\label{eq:qabove}
\end{equation}
by closing the contour above. Similarly one can integrate
\begin{equation}
I_0^b(p) =\frac{1}{2}\int_q  \frac{e^{i (p- q)(L+\epsilon)}}{i(p-q)} \sqrt{\frac{q^2 + 2/\xi^2}{q^2}} \tilde{t}_{n} (q)
\label{eq:qbelow}
\end{equation}
by closing below the real axis. The poles contribution from \eqref{eq:qabove} at $q=p$ simply gives
\begin{equation}
I_0^a(p)_{\mbox{\small poles}} =
\frac{1}{2}  \sqrt{\frac{p^2 + 2/\xi^2}{p^2}} \tilde{t}_{n} (p)
\end{equation}
so that when inserting back in the expression for $a_{mn}$, \eqref{eq:amnL}, we get the contribution from poles
\begin{equation}
	\begin{split}
\left[a_{mn}\right]_{\mbox{\small poles}} =& \, \frac{1}{4} \int_p \tilde{s}_{m}(p) \sqrt{\frac{p^2}{p^2 + 2/\xi^2}}  \sqrt{\frac{p^2 + 2/\xi^2}{p^2}} \tilde{t}_{n} (p) - \frac{1}{4} \delta_{mn}  \\
 =& \, \frac{1}{4} \int_p \tilde{s}_{m}(p)  \tilde{t}_{n} (p) - \frac{1}{4} \delta_{mn} = 0.
\label{eq:amnLpoles}
\end{split}
\end{equation}
This term above would lead to a vanishing entanglement entropy. 

Now we take into account the branch cuts in the integrals \eqref{eq:qabove} and \eqref{eq:qbelow}. To do so we start by rotating $q \rightarrow -iq = y$ and implement this change of variable in both expressions, so that \eqref{eq:qabove} becomes
\begin{equation}
I_0^a(p) = -\frac{1}{4\pi} e^{-ip \epsilon} \int_{i \infty}^{-i \infty} idy\,  \frac{e^{-y\epsilon}}{y+ip} \sqrt{\frac{2/\xi^2 -y^2 }{-y^2}} \tilde{t}_{n} (iy)
\label{eq:qright}
\end{equation}
which now closes to the right, and \eqref{eq:qbelow} is written as
\begin{equation}
I_0^b(p) =\frac{1}{4\pi}e^{i p(L+\epsilon)}\int_{i \infty}^{-i \infty} idy \, \frac{e^{y(L+\epsilon)}}{ip+y} \sqrt{\frac{ 2/\xi^2-y^2}{-y^2}} \tilde{t}_{n} (iy)
\label{eq:qleft}
\end{equation}
which closes to the left. The branch cuts yield respectively the integrals
\begin{equation}
I_0^a(p)_{\mbox{\small bc}} = 
\frac{1}{2\pi} e^{-ip \epsilon} \int_{0}^{\sqrt{2}/\xi} dy\,  \frac{e^{-y\epsilon}}{y+ip}  \frac{\sqrt{2/\xi^2-y^2}}{y}  \tilde{t}_{n} (iy)
\end{equation}
for \eqref{eq:qright} and
\begin{equation}
I_0^b(p)_{\mbox{\small bc}}
=-\frac{1}{2\pi}e^{i p(L+\epsilon)}\int_{0}^{\sqrt{2}/\xi} dy \, \frac{e^{-y(L+\epsilon)}}{ip-y} \frac{\sqrt{2/\xi^2-y^2}}{y} \tilde{t}_{n} (-iy)
\end{equation}
for \eqref{eq:qleft}. The total branch cuts contribution to \eqref{eq:intq} is therefore given by
\begin{equation}
I_0(p)_{\mbox{\small bc}}  = \frac{1}{2\pi}\int_{0}^{\sqrt{2}/\xi} dy \, e^{-y \epsilon} \frac{\sqrt{2/\xi^2-y^2}}{y}  \left[    \frac{e^{-ip \epsilon} }{y+ip}    + (-1)^n \frac{e^{i p(L+\epsilon)}}{ip-y}  \right] \tilde{t}_{n} (iy).
\end{equation}
Here we have used that $-e^{-yL}\tilde{t}_{n} (-iy)=(-1)^n\tilde{t}_{n} (iy)$. By taking the limit $\epsilon \rightarrow 0$ in the above expressions we arrive to
\begin{equation}
	\begin{split}
a_{mn} =& \, \frac{1}{2} \int_p \tilde{s}_{m}(p) \sqrt{\frac{p^2}{p^2 + 2/\xi^2}} I_0(p)_{\mbox{\small bc}}   \\
 =& \, \frac{1}{4\pi} \int_{0}^{\sqrt{2}/\xi} dy \, \frac{\sqrt{2/\xi^2-y^2}}{y}  \int_p \tilde{s}_{m}(p) \sqrt{\frac{p^2}{p^2 + 2/\xi^2}} \\
&\, \times \left[ \frac{1}{y+ip} +(-1)^n \frac{ e^{i pL}}{ip-y} \right] \tilde{t}_{n} (iy).
 	\end{split}
\label{eq:amnLtotal}
\end{equation}

On a next step we calculate 
\begin{equation}
I_1(y) =   \int_p \tilde{s}_{m}(p) \sqrt{\frac{p^2}{p^2 + 2/\xi^2}} \left[  \frac{1}{y+ip} \right]
\end{equation}
and
\begin{equation}
I_2(y) =  \int_p \tilde{s}_{m}(p) \sqrt{\frac{p^2}{p^2 + 2/\xi^2}} \left[ \frac{e^{i pL}}{ip-y} \right]
\end{equation}
first for the case $m=-1$. We see that
\begin{equation}
I_1(y)\big|_{m=-1} =  \int_p \frac{1}{i pL}  \sqrt{\frac{p^2}{p^2 + 2/\xi^2}} \left[  \frac{1 - e^{-ipL}}{y+ip} \right]
\end{equation}
\begin{equation}
I_2(y)\big|_{m=-1} =  \int_p \frac{1}{i pL}  \sqrt{\frac{p^2}{p^2 + 2/\xi^2}} \left[  \frac{e^{ipL}-1}{ip-y} \right] =-I_1(y)\big|_{m=-1}
 \label{eq:int2p}
\end{equation}
have no poles on the real axis, so we integrate  \eqref{eq:int2p} by taking the contour slightly below, and closing above -- note that $y \in (0, \sqrt{2}/\xi)$ guarantees that there are also no poles on the imaginary axis. By rotating $p\rightarrow -ip = x$ we obtain for \eqref{eq:int2p}
\begin{equation}
I_2(y)\big|_{m=-1}
= \frac{1}{2\pi L} \int_{i \infty}^{-i \infty} idx\,  \frac{1}{x} \sqrt{\frac{-x^2}{2/\xi^2-x^2}} \left[ \frac{e^{-xL}-1}{x+y}  \right].
  \label{eq:i1-1}
\end{equation}
In an analogous procedure to the one before, \eqref{eq:i1-1} is integrated to the right of the complex plane to give the branch contribution
\begin{equation}
	\begin{split}
I_2(y)\big|_{m=-1} 
 =& \,    \frac{1}{\pi L}  \int_0^{\sqrt{2}/\xi} \frac{dx}{\sqrt{2/\xi^2-x^2}}  \left[ \frac{e^{-xL}-1}{x+y}  \right].
\end{split}
\end{equation}
In this way one arrives to the matrix row $a_{-1n}$,
\begin{equation}
\begin{split}
a_{-1n}=& \, \frac{1}{4\pi} \int_{0}^{\sqrt{2}/\xi} dy \, \frac{\sqrt{2/\xi^2-y^2}}{y} \left(I_1(y)\big|_{m=-1}-I_2(y)\big|_{m=-1}\right) \tilde{t}_{n} (iy) \\
=& \, \frac{1}{2\pi^2 L} \int_{0}^{\sqrt{2}/\xi} dy  \int_0^{\sqrt{2}/\xi} dx\,   \frac{1}{y}\sqrt{\frac{2/\xi^2 - y^2}{2/\xi^2 - x^2}} \left[ \frac{1-e^{-xL}}{x+y}  \right]   \tilde{t}_{n} (iy) \\
=& \, \frac{1}{2\pi^2 L} \int_{0}^{\sqrt{2}L/\xi} d\bar{y}  \int_0^{\sqrt{2}L/\xi} d\bar{x}\,   \frac{1}{\bar{y}}\sqrt{\frac{2(L/\xi)^2 - \bar{y}^2}{2(L/\xi)^2 - \bar{x}^2}}  \left[ \frac{1-e^{-\bar{x}}}{\bar{x}+\bar{y}}  \right] \tilde{t}_{n} \left(i\frac{\bar{y}}{L}\right),
\label{eq:a-1n}
\end{split}
\end{equation}
setting $\bar{y}=yL$ and $\bar{x}=xL$. 

In a similar manner one calculates for $m=0$ 
\begin{equation}
\begin{split}
I_1(y)\big|_{m=0} =&  \, \int_p \left[\frac{2}{(pL)^2}\left[e^{-ip L}- 1\right] - \frac{1}{ipL} \left[e^{-ip L} + 1\right] \right] \sqrt{\frac{p^2}{p^2 + 2/\xi^2}} \left[  \frac{1}{y+ip} \right] \\
\end{split}
\end{equation}
and
\begin{equation}
I_2(y)\big|_{m=0} = \int_p \left[ \frac{2}{(pL)^2}\left[\frac{1- e^{ipL}}{ip-y}\right] - \left[\frac{1 + e^{i pL}}{ipL(ip-y)}\right] \right]  \sqrt{\frac{p^2}{p^2 + 2/\xi^2}}  =  I_1(y)\big|_{m=0}
\end{equation}
which can again be integrated through a contour closing above the real axis. By rotating $p \rightarrow -ip = x$ the contour closes to the right and captures the branch cut contribution, which gives
\begin{equation}
I_1(y)\big|_{m=0} =   - \frac{1}{\pi L}  \int_0^{\sqrt{2}/\xi} dx\, \left[\frac{2}{xL}\left[\frac{e^{-x L}-1}{x+y}\right] + \left[\frac{1 + e^{-xL}}{x+y}\right] \right]\frac{1}{\sqrt{2/\xi^2-x^2 }}.
\end{equation}
This leads to,
\begin{equation}
\begin{split}	
a_{0n}=& - \frac{1}{2\pi^2 L} \int_{0}^{\sqrt{2}L/\xi} d\bar{y}  \int_0^{\sqrt{2}L/\xi} d\bar{x}\,   \frac{1}{\bar{y}}\sqrt{\frac{2(L/\xi)^2 - \bar{y}^2}{2(L/\xi)^2 - \bar{x}^2}}   \\
& \,\times\left[\frac{2}{\bar{x}}\left[\frac{e^{-\bar{x}}-1}{\bar{x}+\bar{y}}\right]   +  \frac{1 + e^{-\bar{x}}}{\bar{x}+\bar{y}}\right]\tilde{t}_{n}  \left(i\frac{\bar{y}}{L}\right) .
\label{eq:a0n}
\end{split}
\end{equation}

Finally, for $m\geq 1$ we have
\begin{equation}
	\begin{split}
I_1(y)\big|_{m \geq 1} =& \,\frac{1}{L} \int_p \int_0^L ds\, \sin\left(\frac{m\pi s}{L}\right) e^{-ips} \sqrt{\frac{p^2}{p^2 + 2/\xi^2}} \left[  \frac{1}{y+ip} \right] \\
 =& \, \frac{1}{\pi L} \int_0^L ds\, \sin\left(\frac{m\pi s}{L}\right) \int_0^{\sqrt{2}/\xi} dx\,    \frac{x}{\sqrt{2/\xi^2-x^2}} \left[ \frac{e^{-xs}}{x+y} \right]
 	\end{split} 
\end{equation}
and
\begin{equation}
	\begin{split}
I_2(y)\big|_{m \geq 1} =& \,   \frac{1}{L} \int_p \int_0^L ds\, \sin\left(\frac{m\pi s}{L}\right) e^{-ips}  \sqrt{\frac{p^2}{p^2 + 2/\xi^2}} \left[ \frac{e^{i pL}}{ip-y} \right] \\
  =& \, -\frac{1}{\pi L} \int_0^L ds\, \sin\left(\frac{m\pi s}{L}\right)  \int_0^{\sqrt{2}/\xi} dx\,    \frac{x}{\sqrt{2/\xi^2-x^2}} \left[ \frac{e^{x(s-L)} }{x+y} \right],
  \end{split}
\end{equation}
so that for $m \geq 1$,
\begin{equation}
\begin{split}
a_{mn} =& \, \frac{1}{4\pi} \int_{0}^{\sqrt{2}/\xi} dy \, \frac{\sqrt{2/\xi^2-y^2}}{y} \left(I_1(y)\big|_{m \geq 1}+(-1)^nI_2(y)\big|_{m \geq 1}\right) \tilde{t}_{n} (iy) \\
  =& \, \frac{1}{2\pi L} \int_{0}^{\sqrt{2}L/\xi} d\bar{y} \int_0^{\sqrt{2}L/\xi} d\bar{x}\,  \frac{\bar{x}}{\bar{y}(\bar{x}+\bar{y})}\sqrt{\frac{2(L/\xi)^2 - \bar{y}^2}{2(L/\xi)^2 - \bar{x}^2}}  \\
  & \, \times m\left[ \frac{1 - (-1)^m e^{-\bar{x}}}{(m\pi)^2+\bar{x}^2}  \right] \tilde{t}_{n} \left(i\frac{\bar{y}}{L}\right).
\label{eq:amngeq1}
\end{split}
\end{equation}
In all cases $m$ and $n$ have to be of the same parity for $a_{mn}$ not to vanish. 

One can integrate the above expressions for $a_{mn}$ numerically, diagonalize the latter matrix for a chosen (truncated) matrix dimension, and derive the entanglement entropy through \eqref{eq:vNentropy} for an increasing value of $L/\xi$. An important thing to highlight is that the matrix row $a_{m(-1)}$ has divergent elements for all (odd) $m$, as $y \rightarrow 0$ in the integral. This divergence calls for an infrared cutoff $\mu$ to be set by hand, as follows
\begin{equation}
\begin{split}
a_{m(-1)} 
  =& \, \frac{1}{4\pi} \int_{\mu}^{\sqrt{2}L/\xi} d\bar{y} \int_0^{\sqrt{2}L/\xi} d\bar{x}\,  \frac{\bar{x}}{\bar{y}(\bar{x}+\bar{y})}\sqrt{\frac{2(L/\xi)^2 - \bar{y}^2}{2(L/\xi)^2 - \bar{x}^2}}  \\
  & \, \times m\left[ \frac{1 - (-1)^m e^{-\bar{x}}}{(m\pi)^2+\bar{x}^2}  \right] [1+e^{-\bar{y}}];
\label{eq:amngeq1co}
\end{split}
\end{equation}
while all other matrix elements remain finite. Note that $n=-1$ corresponds to a homogeneous mode and the infrared regulator $\mu$ introduced in \eqref{eq:amngeq1co} removes small imaginary momenta $|p|<\mu/L$. For an intuitive argument for the appearance of infrared divergences see section \ref{sec:Phase}.

\subsection{Numerical results}
\label{numerical}

\begin{figure}
	\centering
\includegraphics[width=0.6\textwidth]{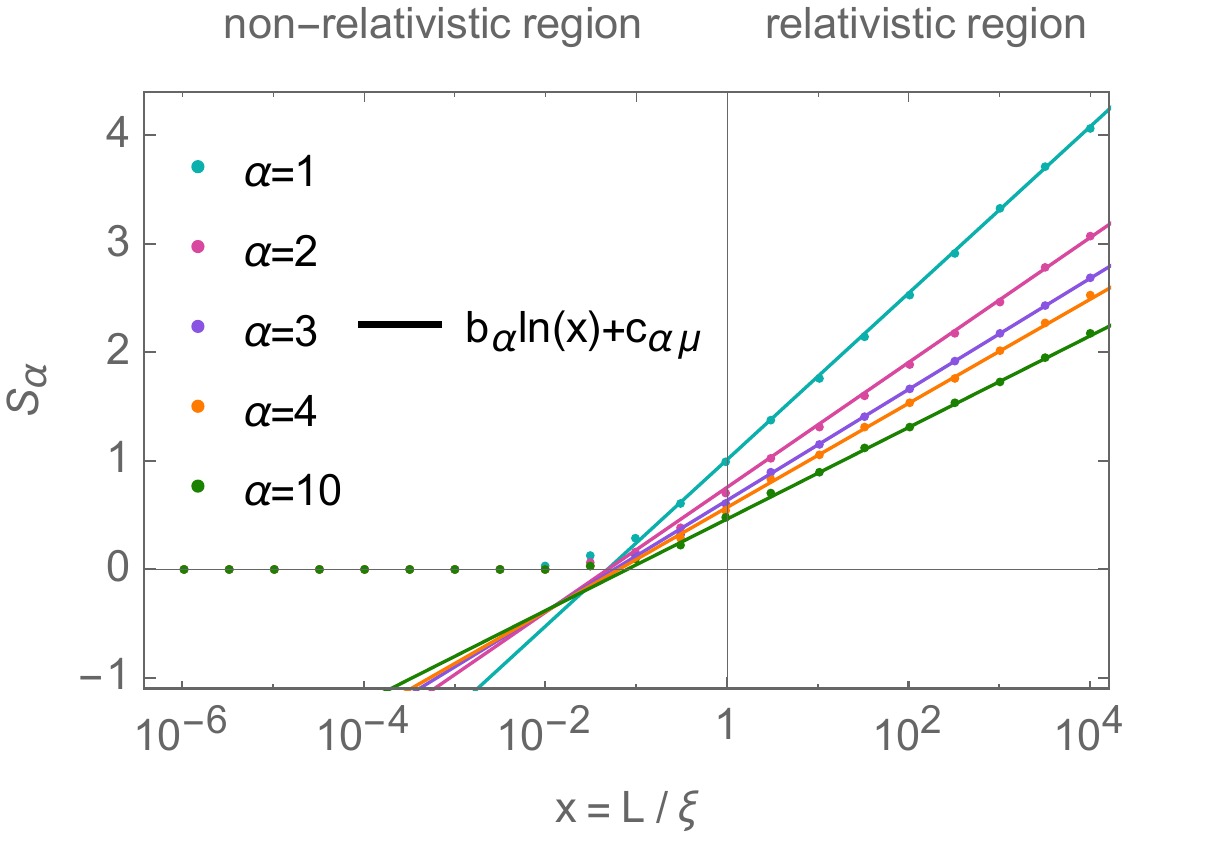}
    \caption{R\'enyi entanglement entropies $S_\alpha$ as a function of the interval length in units of the healing length $x=L/\xi$. We compare different values of $\alpha$, including the von Neumann entanglement entropy $S= S_{\alpha=1}$ as a special case. We find a crossover behavior from vanishing entanglement entropy for small intervals $x=L/\xi \ll 1$ (the ``non-relativistic'' region) to a logarithmic dependence at large $x=L/\xi \gg 1$ (the ``relativistic'' region). For the slope at the relativistic region we recover the result of conformal field theory calculation \cite{CaCa2009, HeRo2016}, see eq.\ \eqref{eq:balphaConformal}. To obtain finite results we have set an infrared regulator $\mu=10^{-5}$ as introduced in eq.\ \eqref{eq:amngeq1co}. Our results were obtained with the numerical scheme described in section \ref{matrix}, through diagonalizing the matrix $a_{mn}$ for a (truncated) matrix dimension $d_M=100$.}
  \label{fig:vnyrenyi}
\end{figure}

In the following we present our numerical results for the entanglement entropy as calculated with the method described above. In Fig.\ \ref{fig:vnyrenyi} we show the R\'enyi entanglement entropy
\begin{equation}
S_\alpha=-\frac{1}{\alpha-1}\ln\mbox{Tr}\{\rho_A^{\alpha}\},
\label{eq:entropycal}
\end{equation}
as a function of $x=L/\xi$, here $\rho_A$ is the reduced density matrix for the interval of length $L$. Besides $\alpha=1$ corresponding to the von Neumann entanglement entropy, we also show the results for $\alpha=2$, $\alpha=3$, $\alpha=4$, and $\alpha=10$. All these results have been obtained from eq.\ \eqref{eq:Rentropy} where the matrix $a$ is evaluated in the Fourier expansion scheme introduced in appendix \ref{ap:modeexpansion}, and truncated to a finite matrix dimension $d_M= 100$. The infrared regulator parameter introduced in eq.\ \eqref{eq:amngeq1co} has been set here to $\mu=10^{-5}$. (The dependence on $d_M$ as well as on $\mu$ will be discussed below.)

Qualitatively one observes in Fig.\ \ref{fig:vnyrenyi} a crossover behavior from a vanishing entanglement entropy $S_\alpha=0$ when the interval is small compared to the healing length $L/\xi \ll 1$, to a logarithmically increasing entanglement entropy for $L/\xi \gg 1$. One may understand this as a crossover from a vacuum-like entanglement entropy as in a non-relativistic quantum field theory (which in fact vanishes) for $L/\xi \ll 1$, to the vacuum-like entanglement entropy in a relativistic situation for $L/\xi \gg 1$.

\begin{figure} 
		\centering
	\includegraphics[width=0.5\textwidth]{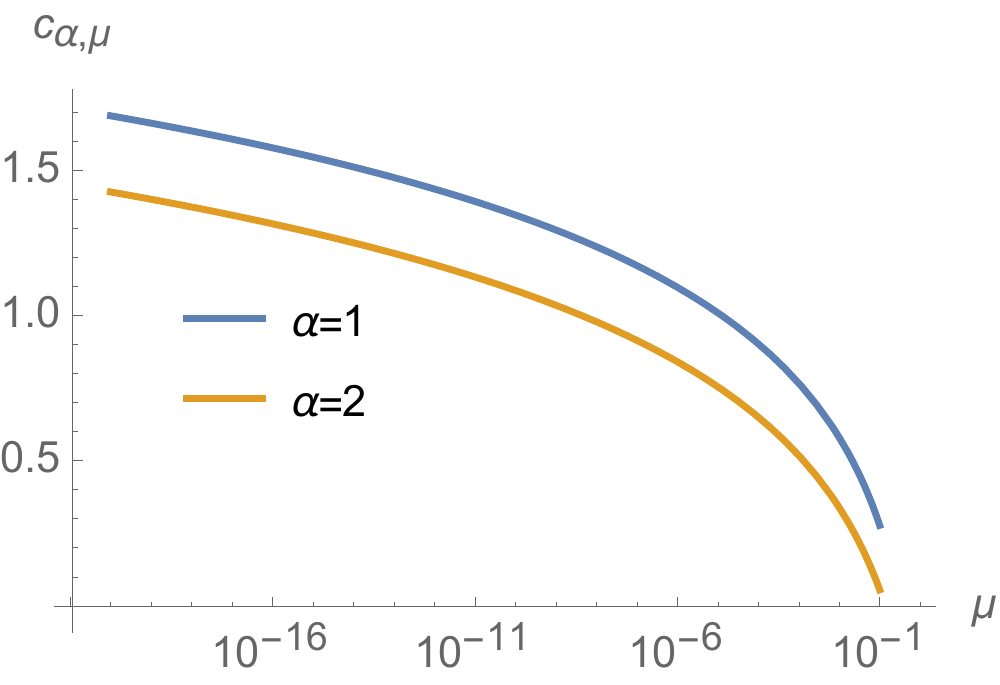}
\caption{Parameter $c_{\alpha, \mu}$ as introduced in \eqref{eq:relentropy} and describing the offset of the entanglement entropy in the relativistic region, as a function of the infrared regulator scale $\mu$. We compare two values for $\alpha$. The dependence of $c_{\alpha, \mu}$ on $\mu$ is rather weak and one finds roughly $c_{\alpha, \mu}\approx 1$ for reasonable values of $\mu$.}
  \label{fig:clambda}
\end{figure}

In the ``relativistic region'' our numerical result is well represented by the behavior
\begin{equation}
S_\alpha \sim b_\alpha \ln (L/\xi) + c_{\alpha \mu},
\label{eq:relentropy}
\end{equation}
where the coefficient $b_\alpha$ matches the result of conformal field theory calculations \cite{CaCa2009}
\begin{equation}
b_\alpha = \frac{c}{6\alpha}(\alpha + 1), 
\label{eq:balphaConformal}
\end{equation}
with a central charge $c=1$. The coefficient $b_\alpha$ is entirely determined by the relativistic regime and independent of the infrared regulator $\mu$. In contrast, the offset parameter $c_{\alpha \mu}$ depends on both, the parameter $\alpha$ and the infrared regulator scale $\mu$ and is shown in Fig.\ \ref{fig:clambda}. From eq.\ \eqref{eq:relentropy} it is clear that the value of $c_{\alpha \mu}$ determines where precisely the crossover from non-relativistic to relativistic entanglement entropy is located. However, from Fig.\ \ref{fig:clambda} one can see that $c_{\alpha \mu}$ is close to unity for reasonable values of $\mu$ so that the transition takes place around $x=L/\xi=e^{-c_{\alpha \mu}/b_\alpha} \approx e^{-1/b_\alpha} \approx 1$.

\begin{figure}
	\centering
	\includegraphics[width=0.5\textwidth]{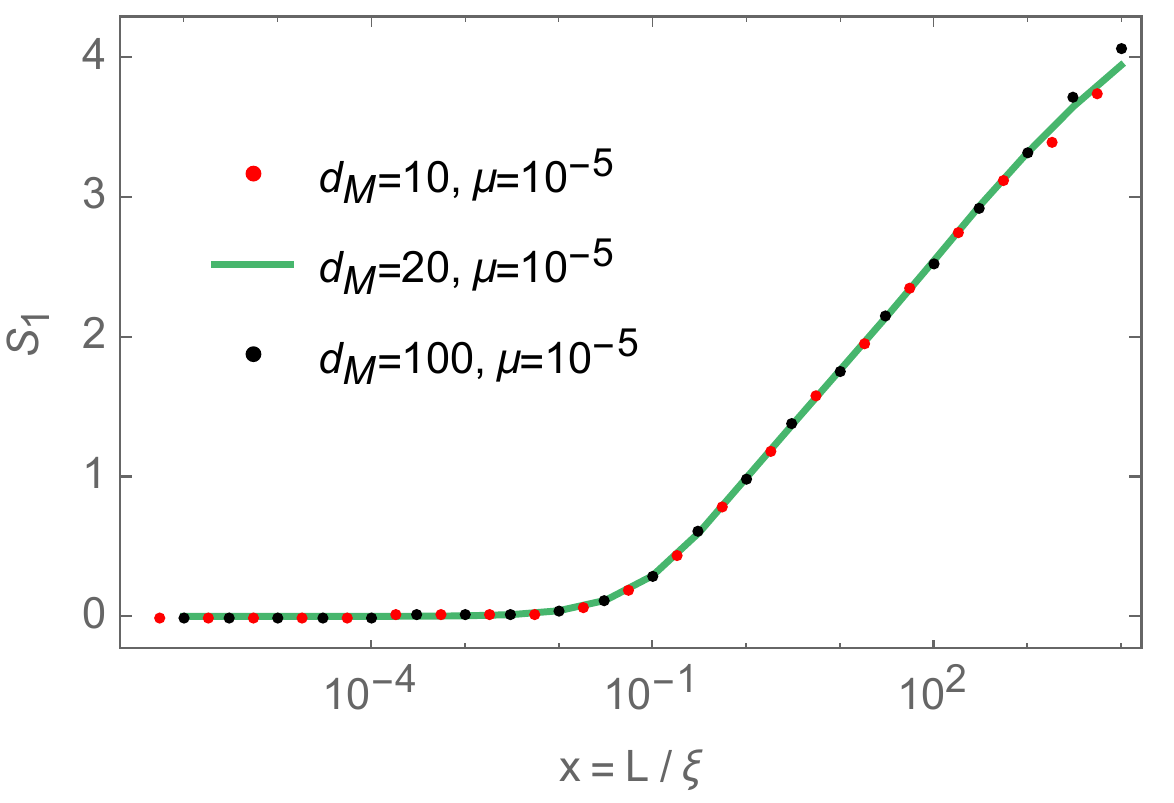}
\caption{Von Neumann entanglement entropy as a function of the interval length in units of the healing length $x=L/\xi$. We chose here $\mu=10^{-5}$ and vary the dependence on the (truncated) matrix dimension $d_M$. The results agree to reasonable accuracy, showing that the numerical scheme introduced in section \ref{matrix} indeed converges in this sense.}
\label{fig:entropydc}
\end{figure}

In Fig.\ \ref{fig:entropydc} we show our numerical result for the von Neumann entanglement entropy $S$ as a function of $x=L/\xi$ for different choices of the (truncated) matrix dimension $d_M$. One can see that the numerical result agrees reasonably well for $d_M=10$, $d_M=20$, and $d_M=100$, which shows that the expansion scheme proposed in eq. \eqref{eq:expansion} and developed in more detail in appendix \ref{ap:modeexpansion} works well and leads to convergent results for the entanglement entropy. For the numerical calculations shown in Figs. \ref{fig:vnyrenyi} and \ref{fig:clambda},  as well as for all further results discussed below, we have fixed $d_M=100$ so that correlation functions are represented by $100 \times 100$ matrices. For the result shown in  Fig. \ref{fig:entropydc} we have set the infrared regulator to $\mu = 10^{-5}$.

\begin{figure}
    \begin{subfigure}{0.5\textwidth}
			\centering
\includegraphics[width=\textwidth]{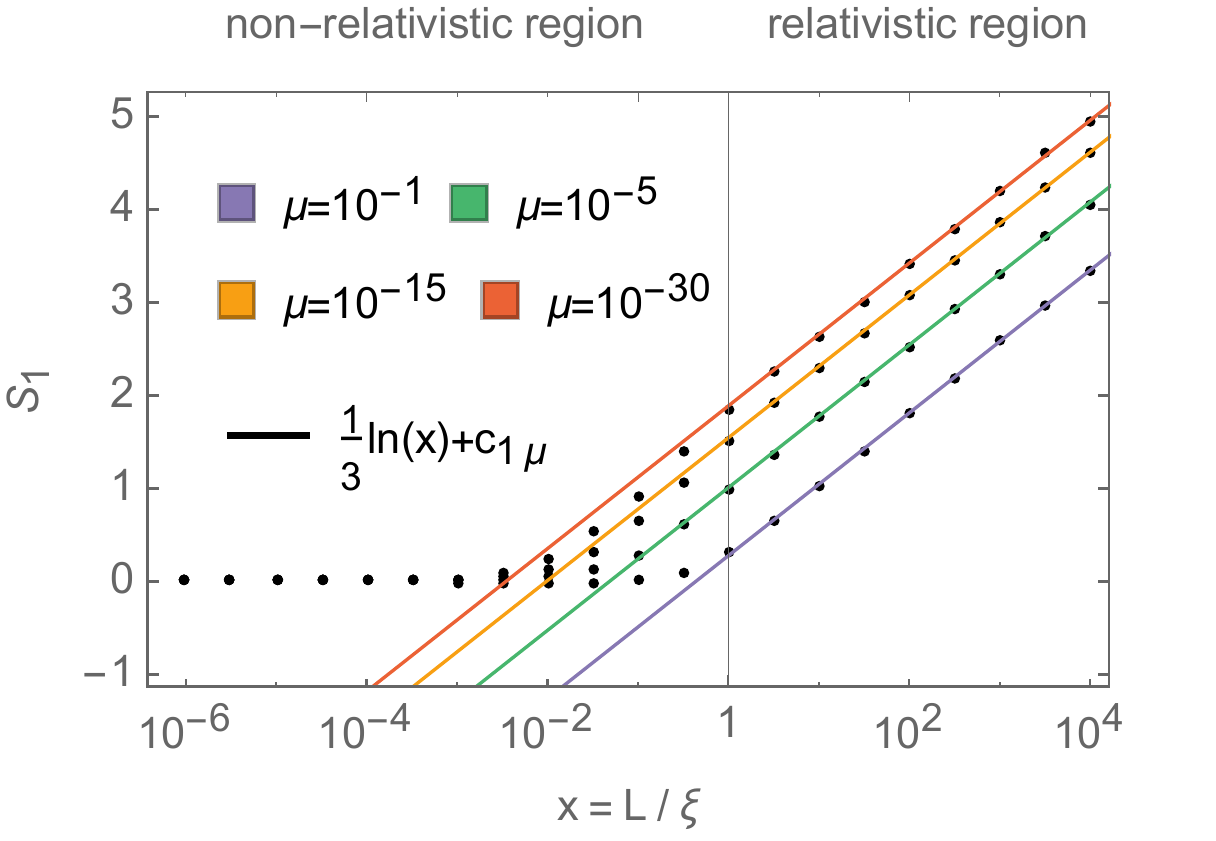}
    \caption{}
    \label{fig:vnentropy}
    \end{subfigure}
    \begin{subfigure}{0.5\textwidth}
			\centering
	\includegraphics[width=\textwidth]{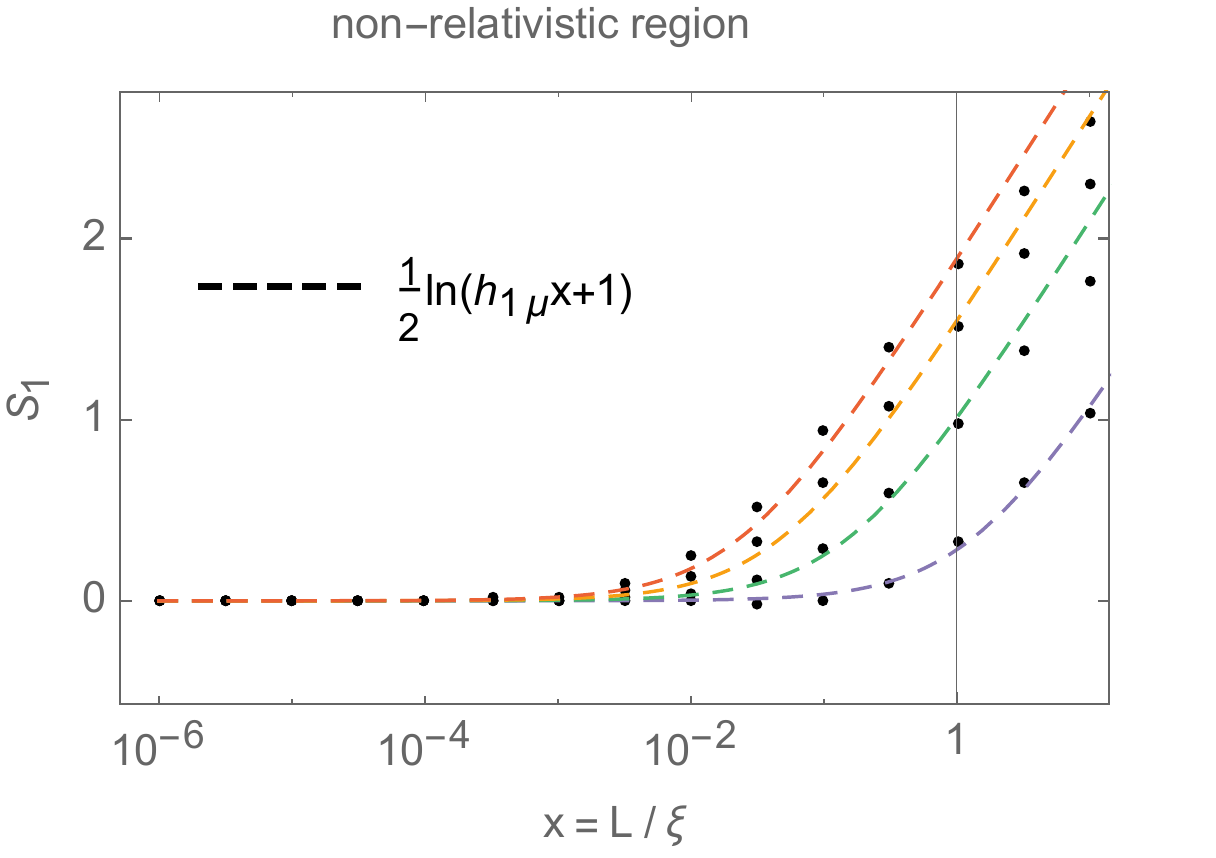}
    \caption{}
    \label{fig:crossover}
    \end{subfigure}
	    \caption{Von Neumann entanglement entropy $S_1$ as a function of the interval length in units of the healing length $x=L/\xi$ for different values of the infrared regulator $\mu$ as introduced in eq.\ \eqref{eq:amngeq1co}. In subfigure (a) on the left we compare the numerical result to an analytic approximation $S_1=\frac{1}{3}\ln(x)+c_{1\mu}$ for the relativistic region, where $c_{1\mu}$ is as shown in Fig.\ \ref{fig:clambda}. In subfigure (b) (on the right), we show the transition region between the non-relativistic and relativistic regimes and compare there to the analytic approximation $S_1=\frac{1}{2}\ln(h_{1\mu}x+1)$, with $h_{1\mu}$ related to $c_{1\mu}$ through eq.\ \eqref{eq:candh}. }	    
 \label{fig:entropylambda}

\end{figure}

It is also interesting to investigate how the entanglement entropy depends on the infrared regulator scale $\mu$. In Fig. \ref{fig:vnentropy} we show the von Neumann entanglement entropy $S_1$ as a function of $x=L/\xi$ for different choices of $\mu$. Interestingly, for increasing $\mu$ one finds that the crossover from non-relativistic vanishing entanglement entropy to a relativistic behavior is moved to larger values of $L/\xi$, but the dependence is only weak. For very small values $\mu \lesssim 10^{-5}$ one can observe that a transition region between non-relativistic and relativistic behavior builds up, which we show in Fig. \ref{fig:crossover}. The functional form in this transition region might be described with reasonable accuracy by
\begin{equation}
S_\alpha \sim \frac{1}{2}\ln(h_{\alpha \mu}x +1),
\label{eq:nrelentropy}
\end{equation}
with $\alpha=1$ in this particular case. The parameter $h_{\alpha \mu}$ is chosen so that the values for the entanglement entropy calculated through \eqref{eq:relentropy} and \eqref{eq:nrelentropy} coincide at $L=\xi$, namely
\begin{equation}
 c_{\alpha\mu}=\frac{1}{2}\ln(h_{\alpha \mu}+1).
\label{eq:candh}
\end{equation}
This gives for instance $h_{1\mu} = -32 \ln(3 \mu)/51$ and $h_{2\mu} = -19 \ln(7 \mu)/51$. The relation \eqref{eq:candh} implies that the entropy on both regions can be fitted by only one free parameter, which depends on the value of the chosen regulator $\mu$.

We also observe that \eqref{eq:nrelentropy} together with $h_{\alpha \mu} \sim - \ln(\mu)$ and $x=L/\xi$ leads to the same dependence on the infrared regulator as in eq.\ \eqref{eq:IRdivergence}. It is therefore likely that one must attribute the behavior of the entanglement entropy in the transition region between the non-relativistic and relativistic regime for $\mu\to 0$ to the entanglement of the homogeneous mode. We should note that the dependence of $S_\alpha$ on $\mu$ is double logarithmic, and therefore so weak that it is unlikely to be of relevance for experiments.

\begin{figure}
			\centering
		\includegraphics[width=0.55\textwidth]{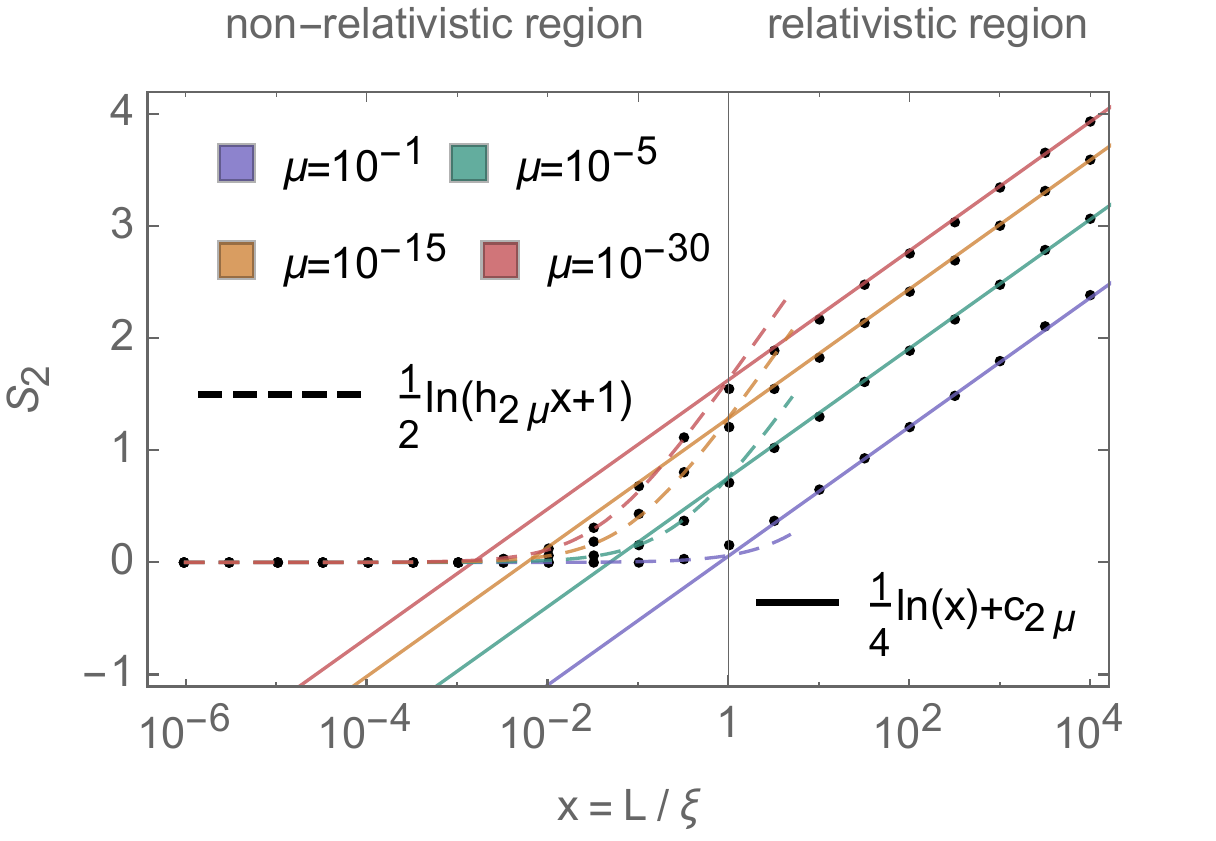}
	    \caption{
		R\'enyi entanglement entropy for $\alpha=2$, for different choices of the infrared regulator $\mu$.
In this case the slope at the relativistic region is $1/4$, in accordance with conformal field theory calculations, see eq.\ \eqref{eq:balphaConformal}. The transition regime between the non-relativistic and relativistic regions can be approximated by eq.\ \eqref{eq:nrelentropy} with parameter $h_{\alpha\mu}$, related to the offset of the entanglement entropy through eq.\ \eqref{eq:candh}. In particular $h_{2\mu}=-19\ln(7\mu)/51$.
	}
    \label{fig:renyi2}
\end{figure}

In Fig.\ \ref{fig:renyi2} we investigate the $\alpha=2$ R\'enyi entanglement entropy and its dependence on the infrared regulator $\mu$. For large $x=L/\xi$ we find here the  behavior $S\sim \ln(x)/4 +c_{2 \mu}$, while for the intermediate $x=L/\xi$ region the ansatz $S\sim \ln(h_{2 \mu}x + 1)/2 $ gives a reasonably good description, at least when $\lambda \lesssim 10^{-5}$. 

Our result for the $\alpha=2$ R\'enyi entanglement entropy can be compared with a calculation for a similar setting using the Lieb-Liniger model \cite{HeRo2016}. More specifically, ref.\ \cite{HeRo2016} has investigated a situation with fixed number of particles in a ring with periodic boundary conditions. This is in contrast to our calculation for fixed chemical potential corresponding at $T=0$ to a coherent state. Nevertheless, for $x=L/\xi \gg 1$ we find the same logarithmic increase of entanglement entropy as ref.\ \cite{HeRo2016} corresponding to the conformal field theory prediction in eqs.\ \eqref{eq:relentropy} and \eqref{eq:balphaConformal}. The differences in the setup explain why the result of ref. \cite{HeRo2016} and ours differ somewhat in the region $L/\xi \ll 1$. While we obtain there $S_2=0$, the result of \cite{HeRo2016} approaches a functional form that can be understood in terms of a quantum mechanical calculation for vanishing interaction $\lambda$ as described in section \ref{QM}. For the transition region to the relativistic regime, as well as in this regime, our result agrees rather well with the one of ref. \cite{HeRo2016} which is based on a more sophisticated approximation to the ground state of interacting bosons in one spatial dimension.

\section{Conclusions}
\label{conclusions}

We have investigated here spatial entanglement in an interacting Bose-Einstein condensate. The result shows an interesting crossover between two limits that are actually both interesting on their own. First, for spatial regions that are small compared to the healing length $\xi =1 /\sqrt{2M\lambda\rho}$ we find that the entanglement entropy approaches zero. This corresponds to the entanglement entropy of the vacuum state for a non-relativistic quantum field theory. That non-relativistic quantum field theories have in their vacuum state a vanishing entanglement entropy is interesting, because that differs greatly from the vacuum state of relativistic quantum field theories, where the entanglement entropy is actually ultraviolet divergent. Second, for regions that are large compared to the healing length, the entanglement entropy for the ground state of an interacting Bose-Einstein condensate behaves in fact like in a relativistic theory, where the sound velocity $c_s$ replaces the velocity of light, and with a natural ultraviolet momentum cutoff given by the inverse healing length $1/\xi$. 

This result is particularly interesting because it shows how one can use the Bose-Einstein condensate for quantum simulations of particular aspects of relativistic quantum field theories. While an experimental investigation of the entanglement entropy seems difficult or out of reach for most fundamental relativistic quantum field theories, the situations is very different for analog systems such as Bose-Einstein condensates of ultra cold quantum gases. Experimental control is already rather good there -- and still getting better. 

The result is also interesting from a theoretical perspective, since it shows how a natural and physical ultraviolet cutoff for the entanglement entropy can actually arise in a quantum field theory that is relativistic in the soft regime. In the short distance or ultraviolet regime of the regularized theory, Lorentz boost symmetry is replaced by Galilean boost symmetry. It might be interesting to explore further whether this insight helps to understand entanglement related problems further, such as the black hole information paradox or the entropy of de Sitter space. A rather interesting perspective is here again that simulations with analog systems are possible, where the space-time metric is replaced by an acoustic metric \cite{Vol2003}.

Other conclusions can be drawn on a more technical level. In particular we have developed a scheme by which the entanglement entropy can be calculated from correlation functions within the region $A$ only. This involves a specifically designed Fourier-type expansion scheme as presented in appendix \ref{ap:modeexpansion}. The information about the exterior enters here in fact only through the boundary conditions, precisely as one would expect. In a next step it might actually be interesting to develop an effective action for the region $A$ and the correlation functions therein, which would have to involve an appropriate boundary action. 

An interesting issue that warrants further study is presented by the infrared divergences of entanglement entropy in the presence of massless particles or gapless excitations in the spectrum. We associate them to entanglement of the homogeneous part of the phase. In general, the infrared divergences are interesting also because they depend on the specific state.

In summary, the spatial entanglement properties of an interacting Bose-Einstein condensate are actually very interesting from a quantum field theoretic point of view, and we are looking forward to further investigations -- both theoretically and experimentally.

\section*{Acknowledgments}
This work is supported by the Deutsche Forschungsgemeinschaft (DFG, German Research Foundation) under Germany's Excellence Strategy EXC 2181/1 - 390900948 (the Heidelberg STRUCTURES Excellence Cluster), by the Deutscher Akademischer Austauschdienst (DAAD, German Academic Exchange Service) under the Länderbezogenes Kooperationsprogramm mit Mexiko: CONACYT
Promotion, 2018 (57437340), SFB 1225 (ISOQUANT) as well as FL 736/3-1.

\appendix
 \section{Fourier transform on a finite interval with arbitrary boundary conditions}
 \label{ap:modeexpansion}
In this appendix we develop a Fourier expansion scheme on a finite interval that does not make any definite assumptions about boundary conditions such as e.\ g.\ periodic, Dirichlet, or von Neumann. This is needed to represent correlation functions on a finite interval efficiently.
 
 \paragraph{Fourier transform on the interval $[0,\pi]$.}
 Let us first consider the finite and closed interval $[0, \pi]$. We want to construct a Fourier-type expansion on this interval that does not assume periodic, Dirichlet or von-Neumann boundary conditions. We explore the ansatz
 \begin{equation}
 	f(z) = f_{-1} + f_{0}\frac{2z - \pi}{\pi} + \sum_{n=1}^\infty f_{n} \sin{(nz)}
 \end{equation}
 which has the following properties.
 \begin{enumerate}
 \item On the boundaries we have $f(0) =  f_{-1} - f_{0}$, $f(\pi) =  f_{-1} +  f_{0}$ and vanishing contributions from $f_n$ with $n \geq 1$.
 \item One may decompose $f(z)=f_{+}(z)+f_{-}(z)$ into a symmetric and an anti-symmetric part with respect to reflections on the point $z= \pi/2$, 
 \begin{equation}
 \begin{split}
 	f_{+}(z) = & f_+(\pi-z) = f_{-1} + \sum_{ n\text{ odd}} f_{n} \sin{(nz)}, \\
 	f_{-}(z) = & -f_{-}(\pi-z) = f_{0}\frac{2z - \pi}{\pi} + \sum_{n \text{ even}} f_{n} \sin{(nz)}.
 \end{split}
 \end{equation}

 \item For the derivatives one has 
 \begin{equation}
 \begin{split}
 	f^\prime(z) = & \frac{2}{\pi} f_{0} + \sum_{n=1}^\infty f_{n} n \cos{(nz)}, \\
 	f^{\prime\prime} (z) = & - \sum_{n=1}^\infty f_{n} n^2 \sin{(nz)},
 \end{split}
 \end{equation}
 and similar for higher oder derivatives.

 \item The expansion coefficients can be obtained from $f(z)$ through the relations
 \begin{equation}
 	f_{-1} = \frac{1}{2}\left[ f(0) + f(\pi)  \right] \quad\quad\quad \text{and} \quad\quad\quad\quad 	f_{0} = \frac{1}{2}\left[- f(0) + f(\pi)  \right],
 \end{equation}
 and for $n \geq 1$
 \begin{equation}
 \begin{split}
 	f_{n} = & \frac{2}{\pi} \int_0^\pi dz \, \left[ f(z) - f_{-1} -f_{0}\frac{2z - \pi}{\pi} \right] \sin (nz) \\
 	= & \frac{2}{\pi}\left[  - \frac{1}{n} f(0)   + \frac{(-1)^n}{n}f(\pi)   + \int_0^\pi dz \, f(z)\sin (nz) \right].
 \end{split}
 \end{equation}
 For odd $n \geq 1$ this simplifies to 
 \begin{equation}
 	f_{n} = \frac{2}{\pi}\left[- \frac{2}{n} f_{-1}  + \int_0^\pi dz \, f(z)\sin (nz)  \right]
 \end{equation}
 while for even $n \geq 2$ one obtains
 \begin{equation}
 	f_{n} = \frac{2}{\pi}\left[\frac{2}{n} f_{0} + \int_0^\pi dz \, f(z)\sin (nz)  \right].
 \end{equation}
 Note that we have a linear relation between the function $f(z)$ in position space and the expansion coefficients $f_n$, which provide a variant of a Fourier space representation.
 \end{enumerate}
 Let us define $s_n(z)$ according to 
 \begin{equation}
 \begin{split}
 s_{-1}(z) = & 1, \\
 s_{0}(z) = & \frac{2z-\pi}{\pi}, \\
 s_{n}(z) = & \sin(nz) \quad\quad\quad \mbox{for} \quad\quad\quad n \geq 1.	
 \end{split} 
 \label{eq:defskernels}
 \end{equation}
 This allows us to write
 \begin{equation}
 	f(z) = \sum_{n=-1}^\infty f_n s_n(z),
 \label{eq:nexp}	
 \end{equation}	
 for $z$ in the interval $[0,\pi]$.  

Similarly, we can define the integration kernels $t_n(z)$ through
 \begin{equation}
 \begin{split}
 t_{-1}(z) = & \frac{\pi}{4}\left[  \delta(z)   + \delta(z-\pi)  \right], \\
 t_{0}(z) = & \frac{\pi}{4}\left[  - \delta(z)   + \delta(z-\pi)  \right], \\
 t_{n}(z) = &  \left[ - \frac{1}{n} \delta(z)   + \frac{(-1)^n}{n}\delta(z-\pi)   + \sin (nz) \right] \quad\quad\quad \text{for} \quad\quad\quad n \geq 1.
 \end{split}
 \label{eq:deftkernels}
\end{equation}
One has then 
 \begin{equation}
 	f_n = \frac{2}{\pi} \int_{0-\epsilon}^{\pi+\epsilon} dz \, f(z) t_n(z).
 \label{eq:xexp}	
 \end{equation}
With \eqref{eq:nexp} and \eqref{eq:xexp} it is now possible to translate between the continuous position or $x$ space and discrete Fourier or $n$ representation.

The $\epsilon$ shift of the boundaries is necessary to make clear that the distributions $\delta(z)$ and $\delta(z+\pi)$ must be included in the integral. Keeping this in mind, we will drop all $\epsilon$'s when no further clarification is important.

\paragraph{Extension to interval $[0,L]$.}
 Let us now extend the expansion developed above to positions $x$ on an interval $[0,L]$ with length $0<L<\infty$. To that end we will use the linear coordinate transformation $x=\frac{L}{\pi}z$. Using the kernels \eqref{eq:defskernels} we write
 \begin{equation}
 \begin{split}
     f(x) = & \frac{1}{L}\sum_{n=-1}^\infty f_n \; s_n\left(\frac{x\pi}{L}\right)\\
     = & \frac{1}{L} \left[ f_{-1} +f_0 \frac{2x-L}{L} + \sum_{n=1}^\infty f_n \sin\left( \frac{n\pi}{L} x \right) \right].
 \end{split}
 \label{eq:nexpL}
 \end{equation}
 The factor $1/L$ has been introduced because this convention has the advantage that for $L\to \infty$ one has the limiting behavior
 \begin{equation}
 \frac{1}{L}\sum_{n=1}^\infty f_n \; \sin\left(\frac{n\pi}{L} x\right) \to 2i \int_0^\infty \frac{dp}{2\pi} \; \tilde f(p) \, \sin(px) = \int_{-\infty}^\infty \frac{dp}{2\pi} \; \tilde f(p) \, e^{ipx},
 \label{eq:limimitingbehaviorLargeL}
 \end{equation}
with $f_n = i \tilde f(p) = -i \tilde f(-p)$ for $p=\frac{n\pi}{L}$. The right hand side of \eqref{eq:limimitingbehaviorLargeL} is of course just the standard Fourier transform. The inverse relation using the kernels \eqref{eq:deftkernels} is given by
 \begin{equation}
 	f_n = \frac{2 L}{\pi} \int_{0}^{\pi} dz \, f\left(\frac{L z}{\pi}\right) t_n(z) = 2 \int_{0}^{L} dx \, f(x) \, t_n\left(\frac{x\pi}{L}\right).
 \label{eq:xexpL}
 \end{equation}
Note that one can alternatively expand a function in position space through the basis $t_n$,
 \begin{equation}
     f(x) = \frac{1}{L}\sum_{n=-1}^\infty \bar f_n \; t_n\left(\frac{x\pi}{L}\right),
 \end{equation}
 with the inverse relation
 \begin{equation}
 	\bar f_n = 2 \int_{0}^{L} dx \, f(x) \, s_n\left(\frac{x\pi}{L}\right).
 \end{equation}

Inserting \eqref{eq:xexpL} in \eqref{eq:nexpL} and {\it vice versa} one finds the completeness and orthogonality relations (for $x,y\in [0,L]$)
 \begin{equation}
 \delta(x-y) =  \frac{2}{L}\sum_{n=-1}^\infty  t_n\left(\frac{y\pi}{L}\right)  s_n\left(\frac{x\pi}{L}\right) =  \frac{2}{L}\sum_{n=-1}^\infty  s_n\left(\frac{y\pi}{L}\right)  t_n\left(\frac{x\pi}{L}\right),
 \label{eq:completeness}
 \end{equation}
 and
 \begin{equation}
 \delta_{mn} = \frac{2}{L} \int_{0}^{L} dx \, s_m\left(\frac{x\pi}{L}\right) \, t_n\left(\frac{x\pi}{L}\right) = \frac{2}{L} \int_{0}^{L} dx \, t_m\left(\frac{x\pi}{L}\right) \, s_n\left(\frac{x\pi}{L}\right).
 \label{eq:orthogonality}
 \end{equation} 
These relations allow to translate chains of operators from position space to the discrete Fourier representation. For example, the trace of an operator can be evaluated in different representations, 
 \begin{equation}
 \begin{split}
 \mbox{tr}\{\mathcal{O}\} = & \, \int_{0}^{L} \int_{0}^{L} dx \,dy \, \mathcal{O}(x,y) \delta(x-y) \\
 = & \, \frac{2}{L} \int_{0}^{L} \int_{0}^{L} dx \,dy \, \mathcal{O}(x,y)  \sum_{n=-1}^{\infty} s_{n}\left(\frac{x\pi}{L}\right)t_{n} \left(\frac{y\pi}{L}\right) = \, \sum_{n=-1}^{\infty}  \mathcal{O}_{nn}, 
 \label{eq:trace}
 \end{split}
 \end{equation}
where we use
\begin{equation}
\mathcal{O}_{mn} =\frac{2}{L} \int_{0}^{L} dx  \int_{0}^{L} dy \, s_{m} \left(\frac{x\pi}{L}\right) \mathcal{O}(x,y) t_{n} \left(\frac{y\pi}{L}\right).
 \label{eq:mnxy}
 \end{equation}
 Alternatively one can also use
 \begin{equation}
 \overline{\mathcal{O}}_{mn} =\frac{2}{L} \int_{0}^{L} dx  \int_{0}^{L} dy \, t_{m} \left(\frac{x\pi}{L}\right) \mathcal{O}(x,y) s_{n} \left(\frac{y\pi}{L}\right)
 \label{mnxy2}
 \end{equation}
 and the operator trace becomes $\text{tr}\{\mathcal{O}\} = \sum_n \overline{\mathcal{O}}_{nn}$.  

\paragraph{Relation to continuous Fourier transform.}
It is useful to relate the discrete Fourier representation \eqref{eq:nexpL} to a standard continuous Fourier transform on the real axis,
 \begin{equation}
 	f(x) =\int_p \, e^{ipx} \tilde{f}(p).
 \end{equation}	
We use here the abbreviation $\int_p = \int_{-\infty}^\infty \frac{d p}{2\pi}$. Specifically, one would like to express the coefficients $f_n$ in terms of $\tilde{f}(p)$,
 \begin{equation}
 	f_n = 2\int_0^{L} dx\, f(x) t_n\left(\frac{x\pi}{L}\right) =\int_p \, \tilde{f}(p) \tilde{t}_n(p) .
 \label{eq:pexp}	
 \end{equation}
 This uses the kernels
 \begin{equation}
 \tilde{t}_n(p)=  2\int_0^{L} dx\, e^{ipx} t_n\left(\frac{x\pi}{L}\right).
 \label{eq:tnp}
 \end{equation}
 Concretely one finds
 \begin{equation}
 \begin{split}
 \tilde{t}_{-1}(p) =& \frac{L}{2}[ 1 + e^{ip L}], \\
 \tilde{t}_{0}(p) =& \frac{L}{2}[  -1 + e^{ip L}  ], \\
 \tilde{t}_{n}(p) =&  \frac{2L}{\pi} \left[  - \frac{1}{n}    + \frac{(-1)^n}{n} e^{ip L} \right]  + 2\int_0^{L} dx\, e^{ipx} \sin\left(\frac{nx\pi}{L}\right) \quad\quad\quad \mbox{for} \quad\quad\quad n \geq 1.
 \end{split}
 \label{eq:ktnp}	
 \end{equation}
 Similarly we use kernels
 \begin{equation}
 	\tilde{s}_n(p) =  \frac{1}{L}\int_{0}^{L} dx \, e^{-ip x} s_n\left(\frac{x\pi}{L}\right).
 \label{eq:snp}
 \end{equation}	
 We can then write
 \begin{equation}
 \begin{split}	
 \mathcal{O}_{mn} =& \, \frac{2}{L} \int_0^L dx\int_0^L dy\, s_m\left(\frac{x\pi}{L}\right)  \mathcal{O}(x,y) \, t_n\left(\frac{y\pi}{L}\right)  \\
 =& \, \frac{2}{L} \int_0^L dx\int_0^L dy \int_p \int_q  s_m\left(\frac{x\pi}{L}\right) e^{-ipx} \, \tilde{\mathcal{O}}(p,q) \, e^{iqy} \, t_n\left(\frac{y\pi}{L}\right) \\
 =& \int_p \int_q  \tilde{s}_m(p)  \tilde{\mathcal{O}} (p,q)  \tilde{t}_n(q).
 \end{split}
 \label{eq:pqtomn}
 \end{equation}
 Concretely we find
 \begin{equation}
 \begin{split}	
 \tilde{s}_{-1}(p) =& \, \frac{1}{ipL} [ e^{ip\epsilon} - e^{-ip(L+\epsilon)}], \\
 \tilde{s}_{0}(p) =& \, \frac{2}{L^2}\left[\frac{e^{-ip (L+\epsilon)}- e^{i p \epsilon }}{p^2}\right] - \left( \frac{L+ 2\epsilon}{L}\right) \frac{1}{ipL}\left[e^{-ip (L+\epsilon)} + e^{i p \epsilon}\right], \\
 \tilde{s}_{n}(p) =& \, \frac{1}{L}\int_{0}^{L} dx \, e^{-ip x} \sin\left(\frac{n x\pi}{L}\right) \quad\quad\quad \mbox{for} \quad\quad\quad n \geq 1.	
 \end{split}
 \label{eq:ksnp}
 \end{equation}
 Furthermore, by combining equations \eqref{eq:orthogonality}, \eqref{eq:tnp} and \eqref{eq:snp} one can see that
 \begin{equation}
 \begin{split}
 	\int_p  \tilde{s}_m(p)\tilde{t}_n(p) = & \, \frac{2}{L} \int_p \int_{0}^{L} dx \, e^{-ip x} s_m\left(\frac{x\pi}{L}\right) \int_{0}^{L} dy \, e^{ip y} t_n\left(\frac{y\pi}{L}\right)  \\
 = & \, \frac{2}{L} \int_{0}^{L} dx \,  s_m\left(\frac{x\pi}{L}\right) t_n\left(\frac{x\pi}{L}\right)  = \delta_{mn},
 \end{split}
 \label{eq:deltamn}
 \end{equation}
 whilst with \eqref{eq:completeness}, \eqref{eq:tnp} and \eqref{eq:snp} one arrives to
 \begin{equation}
 \begin{split}
 	\mathsf{P}_L(p,q) = \sum_{n=-1}^\infty \tilde{t}_n(p) \tilde{s}_n(q) =&\, \frac{2}{L} \sum_{n=-1}^\infty \int_{0}^{L} dx \, e^{ip x} t_n\left(\frac{x\pi}{L}\right) \int_{0}^{L} dy \, e^{-iq y} s_n\left(\frac{y\pi}{L}\right)  \\
 	 =&\,  \int_{0}^{L} dx \,  e^{i(p-q)x} 
 	 =\, \frac{e^{i (p- q)(L+\epsilon)}-e^{-i  (p - q)\epsilon}}{i(p-q)}.
\end{split} 
 \label{eq:sumnpq}	 
 \end{equation}
This last expression can be understood as a projection operator that is unity in the region $(0,L)$ and zero outside, when written in momentum space.

\bibliographystyle{JHEP}
\bibliography{Spatial_entanglement_BEC.bib}

\end{document}